\documentclass[iop]{emulateapj}
\slugcomment{{\sc Submitted to ApJ:} December 20, 2016}
\slugcomment{{\sc Accepted to ApJ:} March 13, 2017}
\usepackage{amssymb,amsmath}
\usepackage{graphicx}
\usepackage{natbib}
\usepackage[colorlinks=true, linkcolor=blue,urlcolor=blue,citecolor=blue]{hyperref}

\begin{document}

\title{MC$^2$: Multi-wavelength and dynamical analysis of the merging galaxy cluster ZwCl 0008.8+5215: An older and less massive Bullet Cluster}

\shorttitle{ZwCl 0008.8+5215 analysis}

\author{Nathan Golovich\altaffilmark{1},
Reinout J. van Weeren\altaffilmark{2}, 
William A. Dawson\altaffilmark{3},
M. James Jee\altaffilmark{1,4},
David Wittman\altaffilmark{1,5}}

\altaffiltext{1}{University of California, One Shields Avenue, Davis, CA 95616, USA}
\altaffiltext{2}{Harvard-Smithsonian Center for Astrophysics, 60 Garden Street, Cambridge, MA 02138, USA}
\altaffiltext{3}{Lawrence Livermore National Laboratory, 7000 East Avenue, Livermore, CA 94550, USA}
\altaffiltext{4}{Department of Astronomy, Yonsei University, 50 Yonsei-ro, Seodaemun-gu, Seoul, South Korea}
\altaffiltext{5}{Instituto de Astrof\'{\i}sica e Ci\^{e}ncias do Espa\c{c}o, Universidade de Lisboa, Lisbon, Portugal}

\email{nrgolovich@ucdavis.edu}

\shortauthors{Golovich et al.}

\label{firstpage}
\begin{abstract}

We present and analyze a rich dataset including Subaru/SuprimeCam, HST/ACS and WFC3, Keck/DEIMOS, Chandra/ACIS-I, and JVLA/C and D array for the merging cluster of galaxies ZwCl 0008.8+5215. With a joint Subaru$+$HST weak gravitational lensing analysis, we identify two dominant subclusters and estimate the masses to be M$_{\text{200}}=\text{5.7}^{+\text{2.8}}_{-\text{1.8}}\times\text{10}^{\text{14}}\,\text{M}_{\odot}$ and $\text{1.2}^{+\text{1.4}}_{-\text{0.6}}\times\text{10}^{\text{14}}\,\text{M}_{\odot}$. We estimate the projected separation between the two subclusters to be $\text{924}^{+\text{243}}_{-\text{206}}\,\text{kpc}$. We perform a clustering analysis of spectroscopically confirmed cluster member galaxies and estimate the line of sight velocity difference between the two subclusters to be $\text{92}\pm\text{164}\,\text{km}\,\text{s}^{-\text{1}}$. We further motivate, discuss, and analyze the merger scenario through an analysis of the 42 ks of Chandra/ACIS-I and JVLA/C and D array polarization data. The X-ray surface brightness profile reveals a merging gas-core reminiscent of the Bullet Cluster. The global X-ray luminosity in the 0.5-7.0 keV band is 1.7$\pm$0.1$\times\text{10}^{\text{44}}$ erg s$^{-\text{1}}$ and the global X-ray temperature is 4.90$\pm$0.13 keV. The radio relics are polarized up to $\text{40}\%$ and along with the masses, velocities, and positions of the two subclusters we input these quantities into a Monte Carlo dynamical analysis and estimate the merger velocity at pericenter to be $\text{1800}^{+\text{400}}_{-\text{300}}\,\text{km}\,\text{s}^{-\text{1}}$. This is a lower-mass version of the Bullet Cluster and therefore may prove useful in testing alternative models of dark matter. We do not find significant offsets between dark matter and galaxies, but the uncertainties are large with the current lensing data. Furthermore, in the east, the BCG is offset from other luminous cluster galaxies, which poses a puzzle for defining dark matter -- galaxy offsets. 

\end{abstract}



\section{Introduction}\label{sec:intro}

Galaxy clusters take shape through a series of hierarchical mergers. Particularly violent mergers are capable of stripping gas off the previously relaxed clusters allowing the approximately collisionless galaxies and dark matter (DM) to run ahead. These mergers are said to be dissociative \citep{Dawson:2012}; examples include the Bullet Cluster \citep{Markevitch04, Clowe06}, the Sausage Cluster \citep{Dawson:2014, Jee:2015}, and several of the Frontier Field clusters \citep[see e.g., ][]{Merten:2011, Golovich:2016}. 

\begin{figure*}[!htb]
\includegraphics[width=\textwidth]{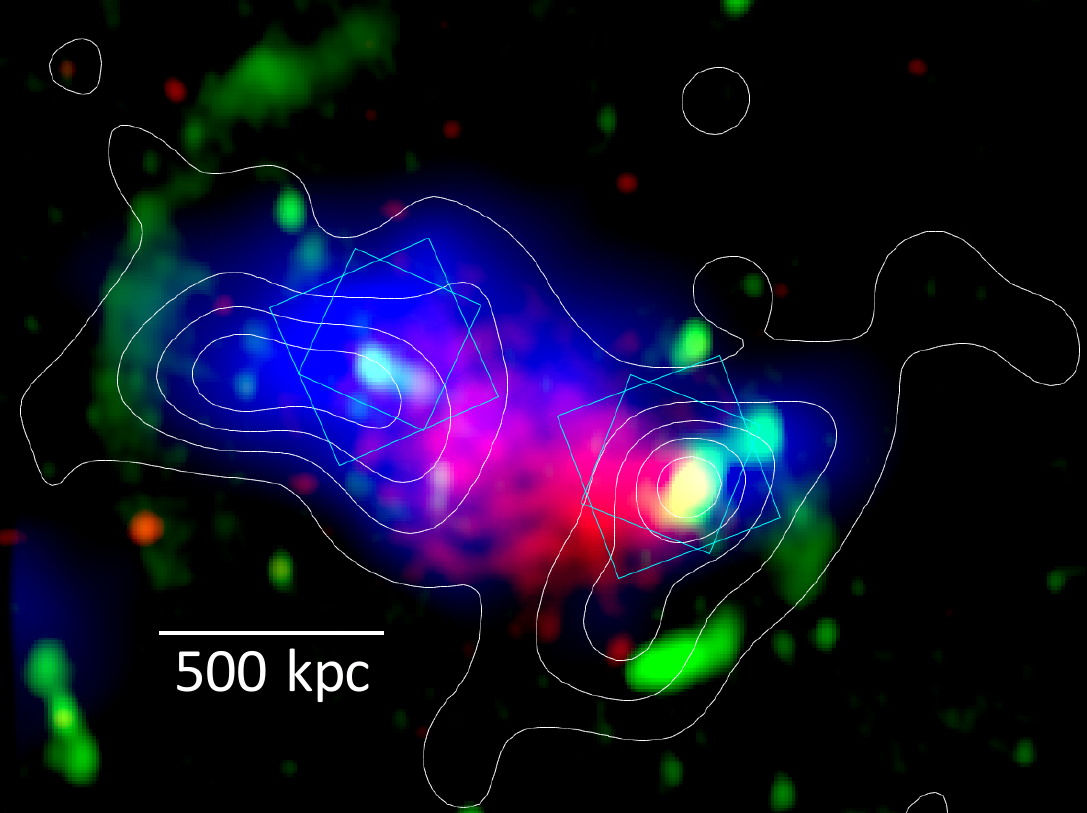}
\caption{Multi-wavelength view of ZwCl 0008 including Chandra X-ray data (red shading), WSRT radio data \citep[green shading;][]{vanWeeren2011}, joint HST--Subaru weak lensing mass (blue shading), and Subaru red sequence luminosity (white contours). The cyan rectangles show the approximate pointings of our HST/ACS+WFC3 observation scheme.}
\label{fig:rgb}
\end{figure*}

Some mergers display shocks in the X-ray emitting gas that are traced by radio relics \citep[e.g.][]{Shimwell:2015, vanWeeren:2017}. When seen edge on, these appear as large (Mpc scale), diffuse radio features \citep[e.g.][]{vanweeren2010, Feretti:2012}. Mergers that occur with collision speeds greater than the intra-cluster medium (ICM) sound speed likely have large scale shocks, but only some have radio relics. The presence of radio relics in a given merging cluster depends on factors that are not directly observable (e.g., magnetic fields in the cluster); however, mergers with radio relics have more tightly constrained dynamical parameters \citep{Ng:2015, Golovich:2016}. This can be due to factors relating to a relationship between the viewing angle of the merger and polarization of the radio relics, and also, the mere presence of a radio relic in a merging cluster seems to imply that the merger axis is near the plane of the sky \citep{Skillman:2013}. ZwCl 0008.8+5215 (hereafter ZwCl 0008, see Figure \ref{fig:rgb}) is a bimodal merger with two radio relics, which enables us to understand and constrain the dynamics of the merger accurately. In this paper we present optical, spectroscopic, X-ray, and radio observations; a wide range of analyses enable us to constrain the dynamics precisely.

\cite{vanWeeren2011} first identified ZwCl 0008 as a double radio relic system while carrying out an extensive search in the 1.4 GHz NVSS, 325 MHz WENSS, and 74 MHz VLSS surveys searching for radio relics in known clusters with possible X-ray emission from the ROSAT All-Sky Survey \citep[RASS,][]{RASS}. For ZwCl 0008, the radio relics were seen first, and cluster emission in RASS corresponding to ZwCl 0008 was subsequently identified, even though it did not meet the criteria for RASS source catalogs. \cite{vanWeeren2011} carried out a radio survey of ZwCl 0008 with Giant-Meterwave Radio Telescope (GMRT) observations at 241 MHz and 640 MHz and Westerbrook Synthesis Radio Telescope (WSRT) observations at 1.3--1.7 GHz in full polarization mode. Two radio relics were identified, with the eastern relic ten times larger than the western relic. Spectral index maps show a steepening trend toward the cluster center for both relics indicating motion away from the center. The spectral indices at the front of the relics were reported to be $-\text{1.2}\pm\text{0.2}\,\,\text{and}\,-\text{1.0}\pm\text{0.15}$ for the east and west relics respectively. Taking these as the injection spectral indices, Mach numbers ($\mathcal{M}$) of $\text{2.2}_{-\text{0.1}}^{+\text{0.2}}$ and $\text{2.4}_{-\text{0.2}}^{+\text{0.4}}$ were reported for the east and west relics. In addition, the polarization was measured at $5-25\%$ for the east relic and $5-10\%$ for the west relic. \cite{vanWeeren2011} also obtained Isaac Newton Telescope (INT)/WFC imaging in V, R, and I bands with 6,000 s exposures. Galaxy isodensity contours suggest a bimodal distribution between the relics. A spectrum of one of the cD galaxies was obtained with a 600 s exposure using William Herschel Telescope (WHT)/ACAM. The spectroscopic redshift was measured to be 0.1032. With this redshift and the RASS count rate, the X-ray luminosity was determined to be $\sim \text{5}\times\text{10}^{\text{43}}\,\text{erg}\,\text{s}^{-\text{1}}$. Using the $L_{X}-T_{X}$ scaling relation from \cite{Pratt:2009}, they found a corresponding temperature of $\sim 4\,\text{keV}$. ZwCl 0008 was studied in a follow up simulation analysis by \cite{Kang:2012}, whose diffusive shock acceleration simulations showed that $\mathcal{M}=\text{2}$ explains the relics in ZwCl 0008 regardless of the level of pre-existing relativistic electrons. They also find a projection angle between 25 and 30$^{\circ}$ to best model the spectral index and radio flux.

Most recently, \citet{Kierdorf:2016} studied ZwCl 0008 with high frequency radio observations at 4.85 GHz and 8.35 GHz with the 100 Effelsberg telescope. They studied the polarization and spectra index of the radio emission and found the polarization fraction of the east relic to vary between 20 and 30$\%$. They find the radio spectrum be 1.44$\pm$0.04, which indicates that ZwCl 0008 is a weak shock. They estimate a Mach number of 2.35$\pm$0.1, which is good agreement with \citet{vanWeeren2011}. The highest frequency radio observations did not cover the west relic, so no estimates are made for this feature. 

In this paper, we add to the understanding of this system with a wealth of data: Subaru/SuprimeCam optical imaging (g and r), a spectroscopic survey with Keck/DEIMOS, 42 ks of Chandra/ACIS-I integration time, six hours of JVLA C and D array observations, and two orbits of HST/ACS+WFC3 optical imaging (F606W and F814W). We will compile the results from analyses of each of these datasets and generate inputs for a Monte Carlo dynamical analysis. In \S\ref{sec:data}, we discuss our observations including target selection, observation, and reduction for each dataset. In \S\ref{sec:products} we generate three galaxy catalogs from our spectroscopic and imaging data to be studied in \S\ref{sec:analysis}, where we analyze the catalogs to estimate the position, mass, and redshift of substructure. In \S\ref{sec:ICM}, we analyze the X-ray and radio data, which will be used in conjunction with the subcluster analysis to develop the merger scenario in \S\ref{sec:scenario}. In \S\ref{sec:mcmac}, we complete a Monte Carlo analysis to constrain the merger dynamics. Finally, in \S\ref{sec:discussion} we discuss and summarize our results. We assume a flat $\Lambda$CDM universe with $H_0 = 70$ $\text{km}$ $\text{s}^{-1}$ $\text{Mpc}^{-1}$, $\Omega_{M} = 0.3$, and $\Omega_{\Lambda} = 0.7$. At the cluster redshift ($z = 0.104$), 1$\arcmin$ corresponds to 115 kpc.

\section{Observations}\label{sec:data}

\subsection{Keck/DEIMOS}\label{subsec:keck}

We conducted a spectroscopic survey of ZwCl 0008 with the DEIMOS \citep{DEIMOS} spectrograph on the Keck II telescope over three separate observing runs (16 January 2013, 14 July 2013 and 5 September 2013). All three observing runs were taken with 1\arcsec wide slits and the 1,200 $\text{line}$ $\text{mm}^{-1}$ grating, tilted to a central wavelength of 6,700 $\text{\AA}$, resulting in a pixel scale of 0.33 $\text{\AA}$ $\text{pixel}^{-1}$, a spectral resolution of $\approx 1$ $\text{\AA}$ (50 $\text{km}$ $\text{s}^{-1}$), and a typical wavelength coverage of 5,400 $\text{\AA}$ to 8,000 $\mathrm{\AA}$. For most cluster member spectra, the wavelength range covered H$\beta$, [O III], Mg I (b), Fe I, Na I (D), [O I], H$\alpha$, and the [N II] and [S II] doublets. This spectral setup enables the study of star formation properties of the cluster galaxies as in \cite{Sobral:2015}. Slits were arranged with a position angle enabling for optimal sky subtraction and to minimize chromatic dispersion by the atmosphere \citep{Filippenko}. We observed a total of four slit masks with approximately 75 slits per mask. For each slit mask, we took three 900 s exposures with the goal of maximizing the number of cluster member spectroscopic redshifts with the survey. 

The Subaru/SuprimeCam imaging was unavailable during spectroscopic survey planning, so we used the INT/WFC imaging described above \citep{vanWeeren2011} to determine the approximate red sequence of the cluster to map out where the galaxies are located. The seeing was 0.9--1.3\arcsec. The low galactic latitude ($b=-9.8647^{\circ}$) and subpar seeing resulted in poor star/galaxy separation in the INT/WFC imaging. We identified a weak red sequence, which was prioritized first. Blue cloud galaxies were targeted with a lower priority in order to fill the mask. We used the DSIMULATOR package to design each slit mask. 

The DEIMOS target selection has some selection effects that will affect the analyses below. First, the $5\arcmin \times 16.7\arcmin$ DEIMOS field of view does not permit us to probe all of the cluster outskirts. This results in some missing data in the spectroscopic survey. Also, multiple slits may not intersect along the dispersion axis of the slit mask, which limits our ability to sample the dense regions of the subcluster centers.

The exposures for each mask were combined using the DEEP2 versions of the $\emph{spec2d}$ and $\emph{spec1d}$ packages \citep{DEEP2:2013}.  $\emph{Spec2d}$ \citep{spec2d} combines the individual exposures, performs wavelength calibration, removes cosmic rays, and performs sky subtraction before generating a processed two and one-dimensional spectrum for each object in a slit. $\emph{Spec1d}$ then fits a template spectral energy distributions (SED) to each 1D spectrum and estimates a redshift using various SED templates for stars, galaxies, and other sources. Finally, we visually inspect the spectra using $\emph{zspec}$ \citep{DEEP2:2013}, assigning quality rankings to each redshift fit \citep[following the convention of][]{DEEP2:2013}. We manually extracted spectra for serendipitous objects that the pipeline missed, and we manually fit redshifts where the pipeline failed to identify the correct fit.

\subsection{Subaru/SuprimeCam}\label{subsec:subaru}

ZwCl 0008 was observed with Subaru/SuprimeCam in two filters. In g, the total integration time was 720 s consisting of four 180 s exposures. In r, the total integration time was 2,880 s consisting of eight 360 s exposures. We rotated the field between each exposure (30$^{\circ}$ for g and 15$^{\circ}$ for r) in order to distribute the bleeding trails and diffraction spikes from bright stars azimuthally to be later removed by median-stacking. This scheme enabled us to maximize the number of detected galaxies, especially for background source galaxies for weak lensing (WL) near stellar halos or diffraction spikes. The median seeing for the g and r images was 0.52$\arcsec$ and 0.57$\arcsec$ respectively.

\begin{figure*}[!htb]
\includegraphics[width=\textwidth]{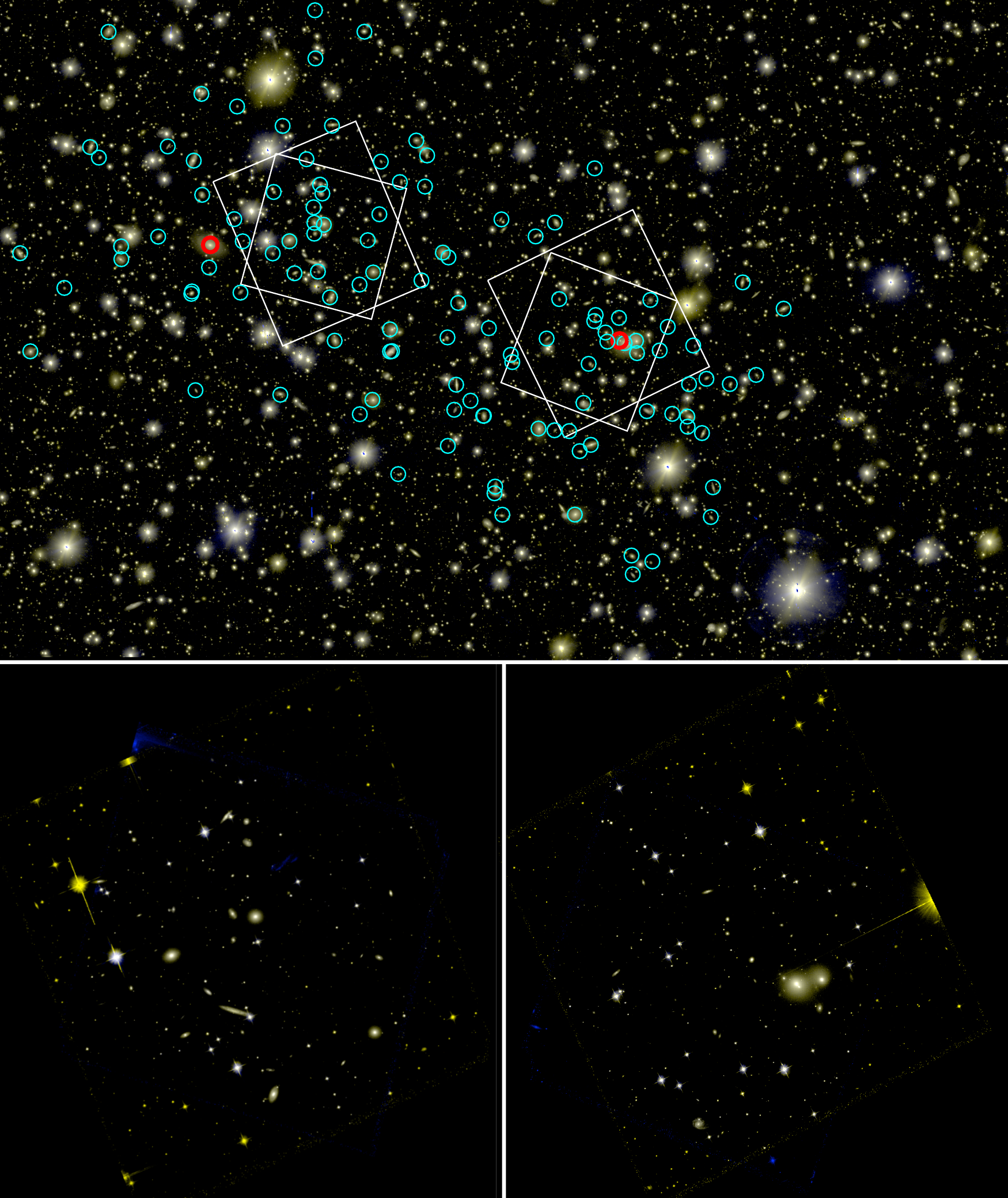}
\caption{\emph{Top:} Subaru SuprimeCam color-composite image with approximate HST ACS and WFC3 fields in white. The field of view approximately matches Figure \ref{fig:rgb}. DEIMOS spectroscopically confirmed cluster members are marked with cyan circles except for the two BCGs that are marked with red circles. The BCGs are too far separated to permit observation of both BCGs with HST in parallel observing mode with ACS and WFC3. \emph{Bottom:} HST color composite images with F606W (WFC3) and F814W (ACS) filters. The eastern pointing is presented in the left panel and the western pointing is presented in the right panel.
\\ }
\label{fig:imaging}
\end{figure*}

The CCD processing (overscan subtraction, flat-fielding, bias correction, initial geometric distortion rectification, etc) was carried out with the SDFRED2 package \citep{Ouchi:2004}. We refined the geometric distortion and World Coordinate System (WCS) information using the SCAMP software \citep{SCAMP}. The Two Micron All Sky Survey \citep[2MASS; ]{2mass} catalog was selected as a reference when the SCAMP software was run. We also rely on SCAMP to calibrate out the sensitivity variations across different frames. For image stacking, we ran the SWARP software \citep{SWARP} using the SCAMP result as input. We first created median mosaic images and then used it to mask out pixels (3$\sigma$ outliers) in individual frames. These masked frames were weight-averaged to generate the final mosaic, which is used for the scientific analysis hereafter. The final image is displayed in the top panel of Figure \ref{fig:imaging} as a color-composite image. Readers are referred to our previous WL analyses for more details on Subaru data reduction \citep[e.g.,][]{Jee:2015,Jee:2016}. 

\subsection{Hubble Space Telescope}\label{subsec:hst}

Two subfields (see Figure \ref{fig:imaging}) of ZwCl 0008 were observed with HST using both Advanced Camera for Surveys (ACS) and Wide Field Camera 3 (WFC3) in parallel during the 2013 October 10 and 2014 January 24 periods under the program HST-GO-13343. Each region was imaged with two orbits of ACS/F814W and two orbits of WFC3/F606W.

Charge transfer inefficiency (CTI) is an important issue when dealing with CCDs in space as high-energy particles damage the detectors and create a number of charge traps for electrons and holes. The effect is severe in both detectors, which if uncorrected for, would leave substantial charge trails and compromise our scientific capability. The current pipeline of the STScI automatically corrects for this effect using the latest pixel based method \citep{HSTCTI}. The importance of such a correction for WL applications is outlined in \cite{Jee:2014}. We use the software MultiDrizzle \citep{multidrizzle} to rectify detector distortions, remove cosmic rays, and create stacks. We used common astronomical objects to measure relative offsets between visits. The estimated alignment error is $\sim$0.01 pixel. We drizzle images with the final pixel scale of 0.05$\arcsec\,\text{pixel}^{-\text{1}}$ and the Lanczos3 kernel. Readers are referred for more details to our previous WL analyses for more details on HST data reduction \citep[e.g., ][]{Jee:2014, Jee:2016}. The combined ACS and WFC3 images are presented in the bottom panels of Figure \ref{fig:imaging}. 

\subsection{Chandra X-ray}

We obtained 42 ks of Chandra/ACIS-I observations of ZwCl 0008 (ObsID: 15318, 17204, 17205) during Cycle 14 and 16. The Chandra data were reduced with the {\tt chav} package following the process described in \cite{Vikhlinin2005}, and using CALDB 4.6.5. The calibration includes the application of gain maps to calibrate photon energies, filtering of counts with ASCA grade 1, 5, or 7 and from bad pixels, and a correction for the position-dependent charge transfer inefficiency. Periods with count rates a factor of 1.2 above the mean count rate in the 6--12 keV band were also removed. Standard blank sky files were used for background subtraction. The final exposure corrected image was made in the 0.5--2.0~keV band using a pixel binning of a factor of four.

\subsection{JVLA Radio}\label{subsec:JVLA}

\begin{table*}[!htb]
\begin{center}
\caption{JVLA Observations}
\bgroup
\def\arraystretch{1.25}
\begin{tabular}{lllll}
&S-band D-array & S-band C-array & \\
\hline
Observation dates &    Oct 19, 2015 &  Oct 9, 2014 \\
Frequencies coverage (GHz)      &  2--4 & 2--4 & \\
On source time per pointing  (hr)   & 1.5 & 0.5    \\
Channel width (MHz) & 2 &  2  \\
Integration time (s)     & 5 & 5 \\
Largest angular scale (arcsec)           &  490 & 490 \\
\end{tabular}
\egroup
\label{tab:jvlaobs}
\end{center}
\end{table*}

ZwCl 0008 was observed with the Jansky Very Large Array in D-array and C-array. All four correlation products were recorded in the 2--4~GHz S-band. Two different pointing centers were observed, one centered on the east relic (00$^{\text{h}}$12$^{\text{m}}$23.60$^{\text{s}}$, 52$^{\circ}$35\arcmin22.00\arcsec) and one on the west relic (00$^{\text{h}}$11$^{\text{m}}$16.50$^{\text{s}}$, 52$^{\circ}$30\arcmin 53.00\arcsec). A summary of the observations is given in Table~\ref{tab:jvlaobs}. 

The data were reduced with the {\tt CASA} software \citep[version 4.5,][]{McMullin:2007}. The data reduction is very similar to that described in \cite{vanWeeren:2016}. In summary, radio frequency interference is flagged in an automatic way employing the \emph{tfcrop} mode of the CASA flagdata task and {\tt AOFlagger}  \citep{Offringa:2010}. We then obtained bandpass, gain, delay, cross-hand delay, polarization leakage and polarization angle solutions for our calibrator sources. These solutions are transferred to the target field. For the target field, the two pointings and array configurations were reduced separately. The calibration solutions were refined via the process of self-calibration. We used w-projection \citep{Cornwell:2005,Cornwell:2008} and MS-MFS clean \cite[{\tt nterms=2},][]{Rau:2011} . 

Clean boxes were used at all stages, created with the {\tt PyBDSM} source detection package \citep{Mohan:2015}. During the self-calibration we employed \cite{briggs_phd} weighting with a robust factor of 0. In the end, the D- and C-array data were combined and imaged together for each pointing. One extra round of phase self-calibration was carried out on the combined data.


\section{Galaxy Catalogs}\label{sec:products}

In this section, we combine our spectroscopic (Keck/DEIMOS, \S \ref{subsec:keck}) and photometric (Subaru/SuprimeCam, \S \ref{subsec:subaru} and HST/ACS and WFC3, \S \ref{subsec:hst}) data to produce galaxy catalogs, which we will use to estimate the subcluster masses, redshifts, and locations, which in turn are the basic inputs for our Monte Carlo dynamical analysis (see \S \ref{sec:mcmac}). 

\subsection{Spectroscopic Catalog}\label{subsec:spectra}

We obtained spectra for 324 objects in the ZwCl 0008 field, of which, 279 objects are assigned a reliable redshift, with the other 45 objects being either too noisy or having too few discernible spectral features to fit a redshift. There are 76 stars and 203 galaxies. Figure \ref{fig:hist} shows the redshift distribution of the 203 high quality \citep[Q$\geq$3, see][]{DEEP2:2013} galaxy spectra (see Table \ref{table:spectra}). Of the galaxies, six are foreground galaxies, 80 are background galaxies, and 117 fall between $0.093 \leq z \leq 0.115$, which is $z_{cluster} \pm \, \text{3000} \, \text{km}\,\text{s}^{-\text{1}}$, where $z_{cluster} = 0.104$. This range is $\sim \pm \text{3}\sigma$, where $\sigma$ is the velocity dispersion. These 117 galaxies are considered cluster members. Since our Keck/DEIMOS spectroscopic survey primarily targeted red sequence cluster galaxies, it is a highly incomplete survey of blue cloud cluster galaxies. Our spectroscopic catalog will be utilized in a Gaussian mixture model (GMM) analysis to test for three dimensional substructure in \S\ref{subsec:subclusters}.

\begin{table*}[!htb]
\begin{center}
\caption{High quality DEIMOS galaxies\footnote{\label{spec_table}Table 2 is published in its entirety in the machine-readable format. A portion is shown here for guidance regarding its form and content.}}
\bgroup
\def\arraystretch{1.25}
\begin{tabular}{llllll}
RA 		 	 & DEC 	     	&	z 	  	& 	$\sigma_{z}$	  	&	Mag (r) 	& 	Comments and spectral features\\
\hline
2.840446	 & 52.52886 	& 0.104259	&	2.49E-05		&	15.8	 	&	BCG west, Mg I (B), Fe I, Na I (D), H$\alpha$\\
3.078217	 & 52.56305 	& 0.105983 	&	1.35E-05		&	16.0	 	&	BCG east, H$\beta$, Mg I (B), Fe I, Na I (D), H$\alpha$, [N II]\\
2.813192	 & 52.50416 	& 0.105324 	&	2.15E-05		&	19.7	 	&	Mg I (B), Na I (D), H$\alpha$ ab\\
2.833348	 & 52.52536 	& 0.097810 	&	2.52E-05		&	18.3	 	&	Mg I (B), Na I (D), H$\alpha$ ab\\
2.829832	 & 52.52540 	& 1.105640  	& 	2.52E-05		&	18.4	 	&	[O II]\\
\end{tabular}\label{table:spectra}
\egroup
\end{center}
\end{table*}

\subsection{Photometric Catalogs}\label{subsec:photometric}

Here we will make use of our Subaru/SuprimeCam and HST/ACS+WFC3 data to generate two catalogs of galaxies to study in various analyses. The first is a red sequence cluster member catalog using the Subaru data. This data will be used to generate galaxy number density maps to study the projected separation between subclusters. The second is a joint-Subaru/HST WL source catalog including shape measurement.

\subsubsection{Red sequence catalog}

We limit the catalogs to a radius of 15$\arcmin$ (1.734 Mpc) from the center of the Subaru field ($\text{RA =}$00$^{h}$11$^{m}$42.4$^{s}$, $\text{DEC =}$52$^{\circ}$31$\arcmin$41.0$\arcsec$). For object detection and shape catalog generation, we refer readers to \cite{Jee:2015} for details, but in brief, we run SExtractor \citep{sextractor} in dual image mode using the r-band image for detection. The blending threshold parameter {\tt BLEND-NTHRESH} is set to 32 with a minimal contact {\tt DEBLEND}$\_${\tt MINCONT} of 10$^{-\text{4}}$. We employ reddening values from \cite{Schlafly:2011} to correct for dust extinction. ZwCl 0008 sits close to the plane of the galaxy ($l=116.7747^{\circ}$, $b=-9.8647^{\circ}$), so the extinction is substantial and variable ($0.7 < \mathrm{A_{g}} < 1.3$ magnitudes, $0.49 < \mathrm{A_{r}} < 0.86$ magnitudes over the Subaru field of view). Finally, we measure object shapes for WL from the r-band images, which provides 0.57$\arcsec$ seeing. The SuprimeCam observations were carried out on the same night as observations of another system (MACS J1752.0+4440) that is covered by the Sloan Digital Sky Survey (SDSS) footprint. We transferred the SDSS zero-point through MACS J1752.0+4440. We observed MACS J1752.0+4440 at an average airmass of 1.12, while we observed ZwCl 0008 at an average airmass of 1.60, so we corrected for the extinction due to the extra 0.48 airmasses. Atmospheric extinction values for Mauna Kea were taken from \cite{Buton:2013}.

\begin{figure}[!htb]
\includegraphics[width=\columnwidth]{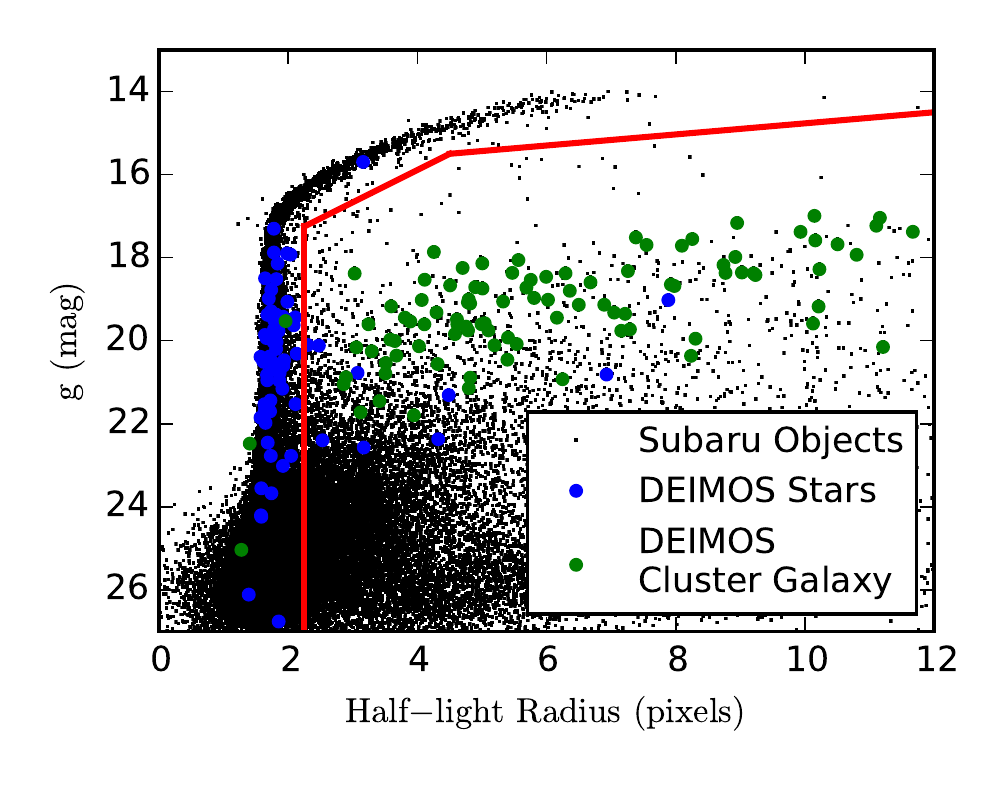}
\caption{Size--magnitude diagram based on dust-corrected Subaru $g$ versus object half-light radius. Overlaid are spectroscopically confirmed cluster members and stars. The red line defines the star--galaxy separation, with galaxies below and to the right of the line.}
\label{fig:sep}
\end{figure}

Since the redshift of ZwCl 0008 is relatively low, it is expected that cluster members will have high signal to noise and correspondingly good photometry. We enforce that objects have uncertainties in their magnitudes of less than 0.5 magnitudes, and we restrict the magnitude range to $14.5<r<22$ and $14.5<g<22$. This eliminates very bright stars that might pass morphological cuts on their size (which is inflated due to saturation and bleeding) as well as false detections at extremely faint magnitudes. The excellent seeing ($0.57\arcsec$) of the Subaru r-band imaging enables accurate star-galaxy separation via cuts on the half-light radius (see Figure \ref{fig:sep}). We eliminate objects with a half-light radius of less than 2.25 pixels ($0.45\arcsec$). The rest of the boundary changes slope with the stellar track. Over 50$\%$ of the objects in the Subaru catalog are removed. The fraction of stars is high because of the low galactic latitude. There are a high number of binary stars and blended objects where the dominant light is from a star which explains the blue points to the right of the stellar track in Figure \ref{fig:sep}. 

\begin{figure}[!htb]
\includegraphics[width=\columnwidth]{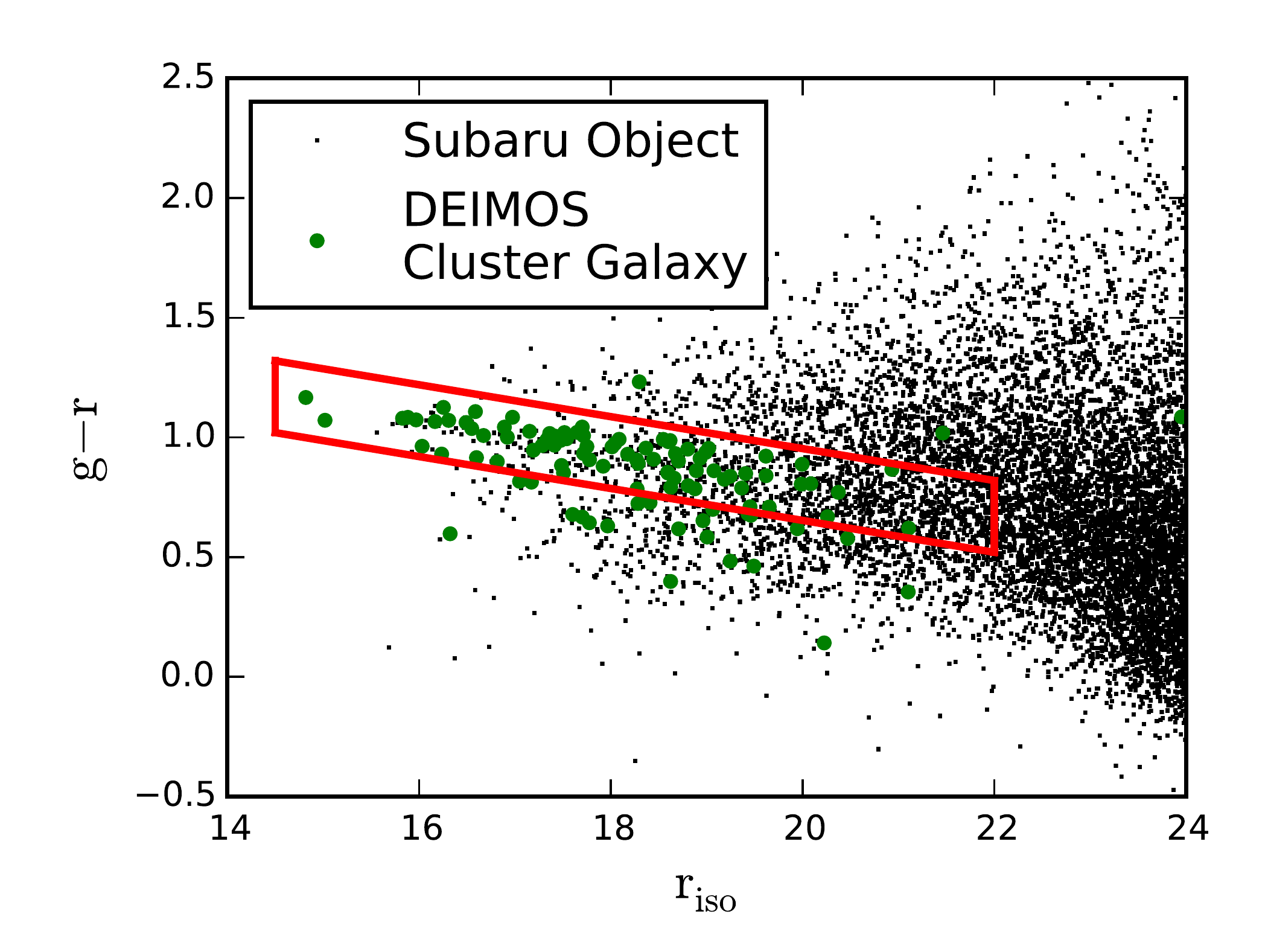}
\caption{Color-magnitude diagram based on dust-corrected Subaru $\text{g}-\text{r}$ versus $\text{r}$-isophotal magnitudes. Overlaid are the 117 spectroscopically-confirmed cluster members. These lie along the red sequence because they were largely targeted via a red sequence selection as described in \S \ref{subsec:keck}. The two BCGs are the left-most green dots in the red boxed region.}
\label{fig:redsequence}
\end{figure}

A clear and tight red sequence is visible in a color magnitude diagram, which is presented in Figure \ref{fig:redsequence}. We highlight this red sequence by plotting matched spectroscopically confirmed cluster members (primarily selected from the red sequence in INT/WFC imaging). Some spectroscopic members are below the red sequence box. These are blue cloud galaxies of the cluster, which were significantly under sampled. In the 15$\arcmin$ radius field, 19,014 galaxies are left after the star-galaxy separation (27 galaxies arcmin$^{-\text{2}}$).

After applying the red sequence cut from Figure \ref{fig:redsequence}, we find 950 cluster member galaxies within a 15$\arcmin$ radius. We estimate the purity of this sample of red sequence galaxies by considering the population of spectroscopic stars and galaxies in the red sequence selection. These should be considered rough estimates, as the red sequence is defined out to $r=22$, while the spectroscopic sample's completeness significantly decreases at fainter magnitudes. Within the prescribed red sequence selection region, there are 101 objects with secure redshifts. Of these, 77 are cluster members; zero are foreground galaxies; 20 are background galaxies; four are stars. We mapped the contaminants for spatial significance, and there is no obvious projected structure; thus, we do not expect significant bias in the subcluster location estimates from the red sequence sample. We will make use of this catalog in \S \ref{subsec:2d} to estimate the projected distance between subclusters and the projected offset between cluster components.  

\subsubsection{Weak lensing source catalog}\label{sssec:sourcecat}

To generate the lensing source catalog we combine the Subaru and HST data. The goal is to compile a catalog of background galaxies with shape and color measured for each galaxy. 

For the Subaru data, we rely on color-magnitude relations to differentiate between cluster members and lensing sources. The color is defined by $g-r$, which brackets the 4000$\AA$ break at the redshift of the cluster. For the Subaru catalog, we make use of the red sequence (see Figure \ref{fig:redsequence}) to eliminate likely cluster members from the source catalog. We allow objects redder than the red sequence at any magnitude, as well as any objects bluer than the red sequence and fainter than $r=22$. In addition to the color and magnitude selection, we apply the following shape criteria: the post-seeing half light radius must be greater than 0.44$\arcsec$, the shape uncertainty must be less than 0.3 after PSF deconvolution, and the semi-minor axis must be greater than 0.3. 

For the HST catalog objects are selected if their magnitude in F814W is between 20 and 27, their ellipticity error is less than 0.25, their half life radius is greater than 1.3 pixels, and their semi-minor axis is greater than 0.4 pixels. This results in a source density of  76 and 82 arcmin$^{-\text{2}}$ for the east and west pointing respectively. 

The photometric coverage is characterized by three regions. For the Subaru-only and the HST region with only ACS/F814W is available, we use Subaru colors. To estimate the source redshift distribution, we make use of the photometric catalog from \cite{Dahlen:2010} from the Great Observatories Origins Deep Survey \citep[GOODS;][]{GOODS}. Specifically, we used the GOODS-S catalog, which covers $\sim$160 arcmin$^{\text{2}}$. The three photometric regions need to be estimated separately. For the first region, with Subaru-only photometry, we perform a photometric transformation of the g--r color to match the ACS colors. For each region, we estimate the angular diameter distance ratio: $\beta=D_{ls}/D_{s}$, where $D_{ls}$ and $D_{s}$ are the angular diameter distances between the lens and the source and between the observer and the source, respectively. Knowledge of $\beta$ is required to estimate the surface mass density. For the Subaru-only region, we estimate $\bar{\beta}=\text{0.817}$, and after correcting for the difference in depth of the GOODS survey with our data, we determine $\bar{\beta}=\text{0.805}$. For the ACS only region, we find $\bar{\beta}=\text{0.867}$. Finally, for the ACS+WFC3+Subaru region, we assume F814W to match F775W in GOODS and find $\bar{\beta}=\text{0.851}$.


\section{Galaxy and Mass Analysis}\label{sec:analysis}

In this section, we analyze the three galaxy catalogs developed in \S \ref{sec:products}. Our goal is to identify the merging subclusters and estimate their redshifts, masses, and projected separations. These quantities will be input into the dynamical analysis of \S\ref{sec:mcmac}. 

\subsection{Subcluster Analysis}\label{subsec:subclusters}

We have generated two catalogs of cluster member galaxies. The spectroscopic catalog is pure but highly incomplete, and it is limited by the spectroscopic selection effects. The red sequence catalog was trained by the location of the spectroscopic sample and is more complete, but it contains contaminants. A number of one dimensional (velocity), two dimensional (projected space), and three dimensional (velocity + projected space) clustering algorithms have been developed to test for clustering within discrete data \citep[see][for a review]{Pinkney:1996}. Our primary goal is to determine the number and location of subclusters that are dynamically bound and active participants in the merger. That being said, it is helpful to identify any foreground or background clustering to rule out groups of galaxies from the merging event \citep[see e.g., Abell 781;][]{Wittman06}. Furthermore, any line of sight (LOS) structure must also be accounted for and carefully modeled to properly infer the mass distribution of a cluster. 

\begin{figure}
\includegraphics[width=\columnwidth]{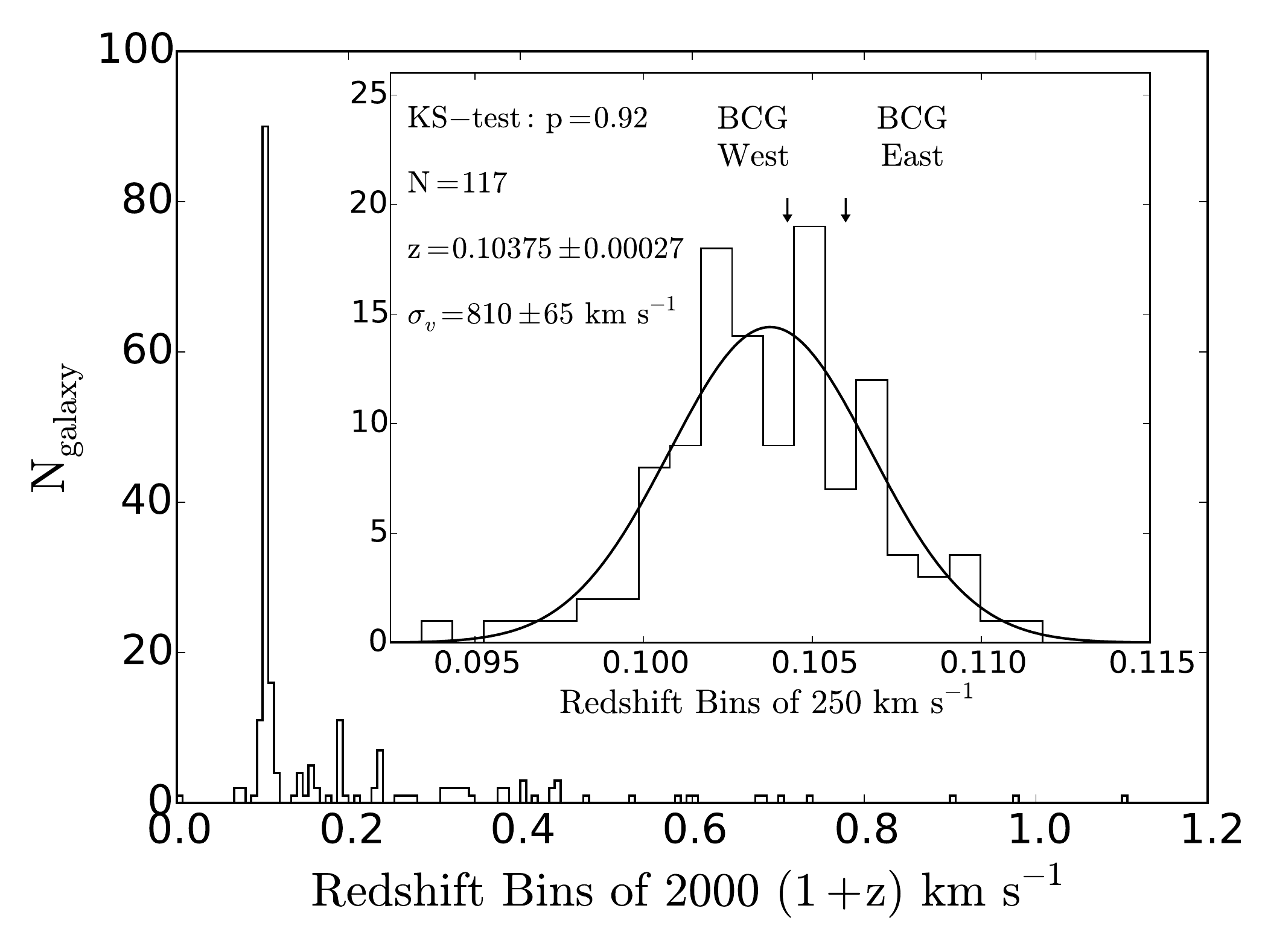} 
\caption{\emph{Main:} Redshift distribution of the 203 Keck DEIMOS high quality ($\text{Q}\ge\text{3}$) galaxy spectroscopic redshifts. The over-density near $\text{z}=\text{0.103}$ is clear and corresponds to the 117 spectroscopically confirmed members of ZwCl 0008. Also shown are the five foreground and 81 background galaxies with high quality redshifts. \emph{Inset:} A zoom in of the spectroscopic histogram near the cluster redshift. The east and west subcluster brightest cluster galaxy (BCG) redshifts are indicated with arrows.}
\label{fig:hist}
\end{figure}

Figure \ref{fig:hist} shows the redshift distribution from the spectroscopic survey. The redshift distribution of ZwCl 0008 is well modeled by a single Gaussian (p-value of 0.959). For this reason, we will forgo detailed one-dimensional analyses. The presence of radio relics and bimodal galaxy distributions revealed by \cite{vanWeeren2011} provides sufficient evidence that ZwCl 0008 is in a merging state and composed of more than one subcluster. We can infer that the subclusters must not have a substantial LOS velocity difference in the observed state because the redshift distribution is well modeled by a single Gaussian. This leaves the possibility that the subclusters are either moving in the plane of the sky and/or near apocenter.

\begin{figure}[!htb]
\includegraphics[width=\columnwidth]{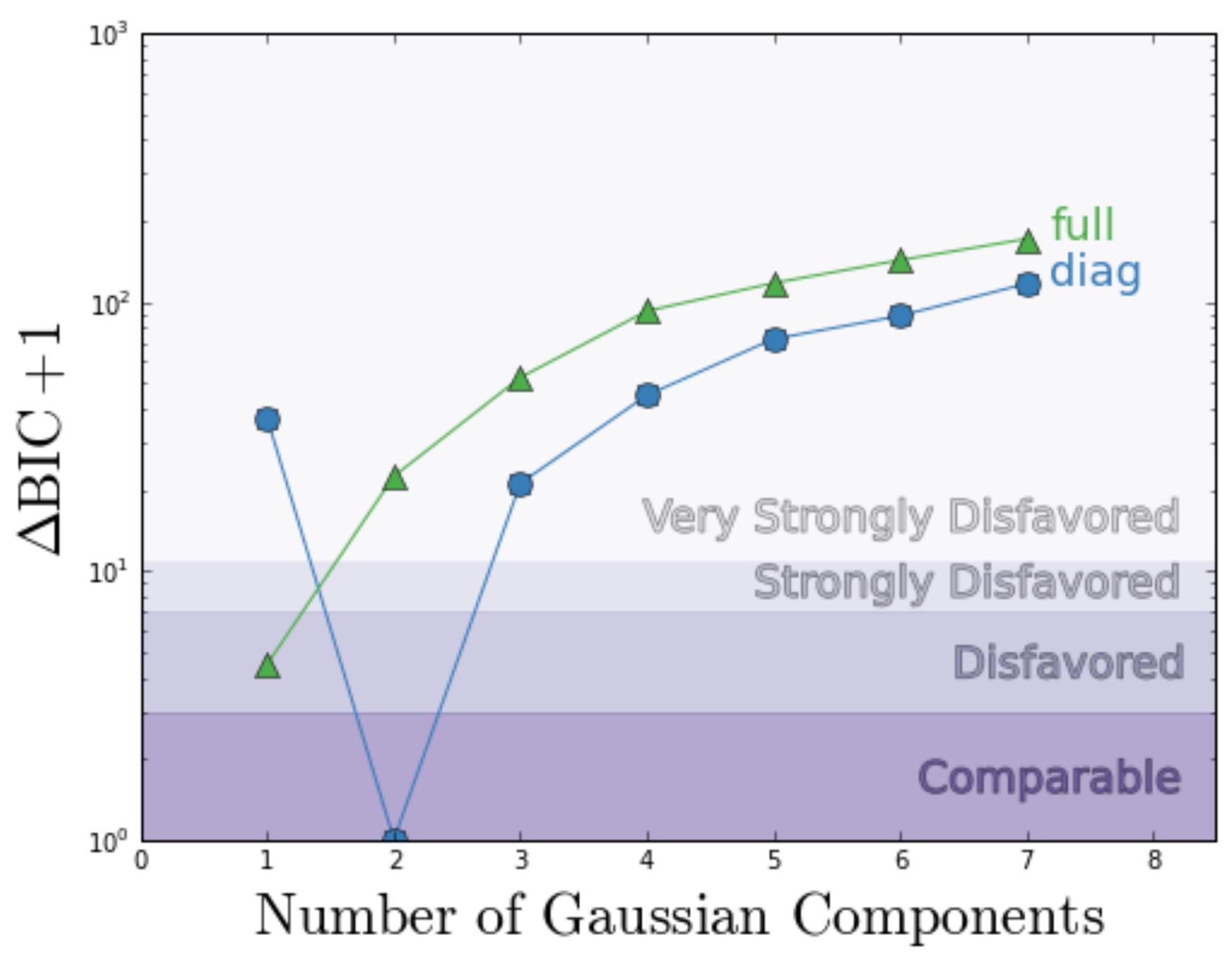}
\caption{$\Delta$BIC plot comparing GMM fits to the three-dimensional (right ascension, declination, and redshift) distribution of all the cluster member spectroscopic galaxies with varying number of Gaussian components and covariance type. We plot the results for models with \emph{diag} (blue circles) and \emph{full} (green triangles) covariance types. The purple shaded regions roughly denote how a given model compares with the model with the lowest BIC score \citep{Kass:1995}. The most economical fit is a two-component model with \emph{diag} covariance structure.}
\label{fig:gmmselect}
\end{figure}

To utilize the information of all three dimensions in the spectroscopic catalog, we use a GMM algorithm from the python package SciKit-Learn \citep{sklearn}. The first implementation of the following method was presented in \cite{Dawson:2014}. This python package gives options to the form of the covariance matrix. We vary the number of components (in both the full and diagonal covariance models), and in Figure \ref{fig:gmmselect} we compare the models by their Bayesian information criterion (BIC) relative to the lowest BIC amongst the models tested. A two component, diagonal model is preferred for the spectroscopic catalog. A one component full covariance model is also an acceptable fit; however, the model is highly elliptical and we deemed it unphysical.

\begin{figure*}[!htb]
\begin{center}
\includegraphics[width=\textwidth]{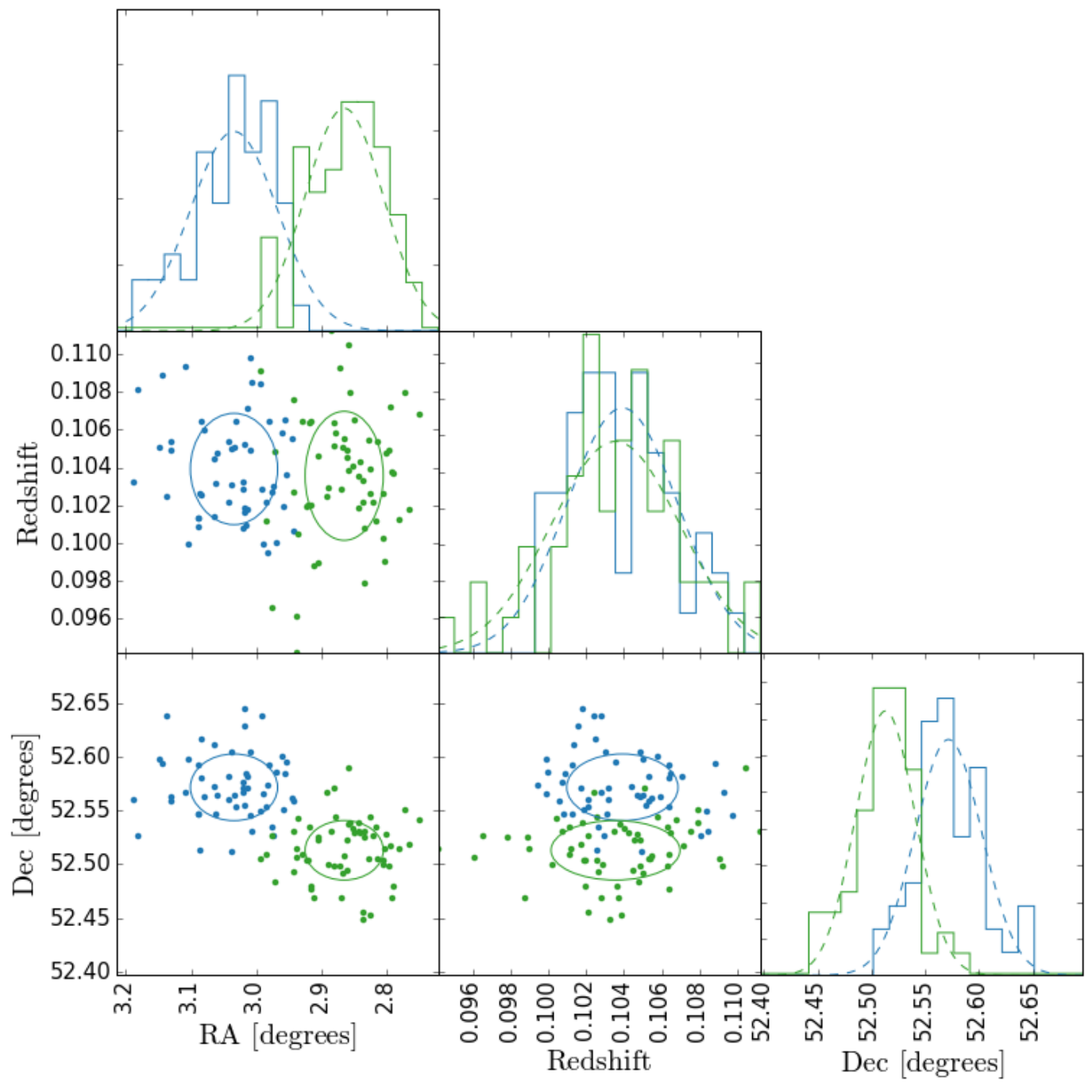}
\caption{The three-dimensional distribution of the spectroscopic cluster members and their most likely subcluster membership assignment for the two-dimensional \emph{diag} model from Figures 9 and 10. For the projected one-dimensional distributions, we plot the marginalized Gaussian components for the best fit model with dashed lines. For the projected two-dimensional distributions, we plot marginalized 68\% confidence ellipses of the best fit model Gaussian components.}
\label{fig:gmmresult}
\end{center}
\end{figure*}

We use this model to infer membership of the individual galaxies in each subcluster. Galaxies are assigned to the Gaussian that gives the largest probability of membership.  A corner plot is presented in Figure \ref{fig:gmmresult} detailing the model. The ellipses in the scatter plots show the two-dimensional projections of the three-dimensional Gaussians that were determined to explain the spectroscopic catalog. The one dimensional histogram panels show normalized right ascension, declination, and redshift histograms of the determined subclusters. In the bottom-left scatter plot, projected locations of the two subclusters are shown. The GMM has largely identified the two subclusters to be split down the middle of the spectroscopic survey in projected space.

\subsection{Subcluster Redshifts}\label{subsec:redshifts}

As mentioned, one of the three primary inputs for the dynamical analysis of \S \ref{sec:mcmac} is an estimate of the LOS velocity difference between the merging components. In the previous subsection, we have identified two merging subclusters, and here we will analyze their redshift distribution. We estimate the redshift and velocity dispersion of the two subclusters by making use of the biweight statistic and bias-corrected 68$\%$ confidence limits \citep{Beers1990} applied to bootstrap samples of each subcluster's spectroscopic redshift catalog. We find very similar redshifts for the two subclusters, which have a relative LOS velocity difference of just $82\,\pm\,150\,\text{km}\,\text{s}^{-\text{1}}$. The velocity dispersions of the two subclusters ($\text{736}^{+\text{76}}_{-\text{50}}\,\text{km}\,\text{s}^{-\text{1}}$ and $\text{895}^{+\text{117}}_{-\text{93}}\,\text{km}\,\text{s}^{-\text{1}}$ for the east and west subcluster, respectively) can be used to estimate the mass of the system \citep{Evrard:2008}. We find the west subcluster to be $\text{7.7}^{+\text{3.4}}_{-\text{2.1}}\,\times \text{10}^{\text14} \text{M}_{\odot}$ and the east subcluster to be $\text{4.3}^{+\text{1.5}}_{-\text{0.8}}\,\times \text{10}^{\text14} \text{M}_{\odot}$. Velocity dispersion mass estimates have been shown to be biased high in disturbed systems due to the overlap of the two potential wells increasing the velocities of galaxies \citep{Carlberg:1994}. However, \citet{Pinkney:1996} demonstrated that the bias is substantially diminished by the time the subclusters have reached apocenter $\sim$1 Gyr after pericenter in a 3:1 mass ratio simulation, which is similar to the configuration of ZwCl 0008. We present these masses for comparison to the WL mass estimates of \S\ref{subsec:masses}. 

There is bias introduced by the GMM's inability to properly assign membership in the areas where the two overlap. To test this effect, we simulated the cluster repeatedly by randomly selecting galaxies from 3D Gaussians with a grid of values near the values from the heretofore analysis. We cut the data to the DEIMOS footprint, and ran the simulated data through the two-component 3D GMM. We found a systematic bias in the measured velocity information. The velocity difference is underestimated by $\text{10}\pm\text{68}\,\text{km}\,\text{s}^{-\text{1}}$. This is to be expected because contaminants tend to decrease (increase) the inferred line of sight velocity of the higher (lower) redshift subcluster. Additionally, the subcluster velocity dispersions are found to have been artificially inflated by $\text{5.1}\pm\text{70}\,\text{km}\,\text{s}^{-\text{1}}$ and $\text{7.0}\pm\text{85}\,\text{km}\,\text{s}^{-\text{1}}$ for the east and west subclusters, respectively. The bias is small because the two subclusters have very similar line of sight velocities and unsubstantial differences in velocity dispersion to begin with; however, this effect is an important consideration in general for any subclustering algorithm. 

\begin{figure}[!htb]
\includegraphics[width=\columnwidth]{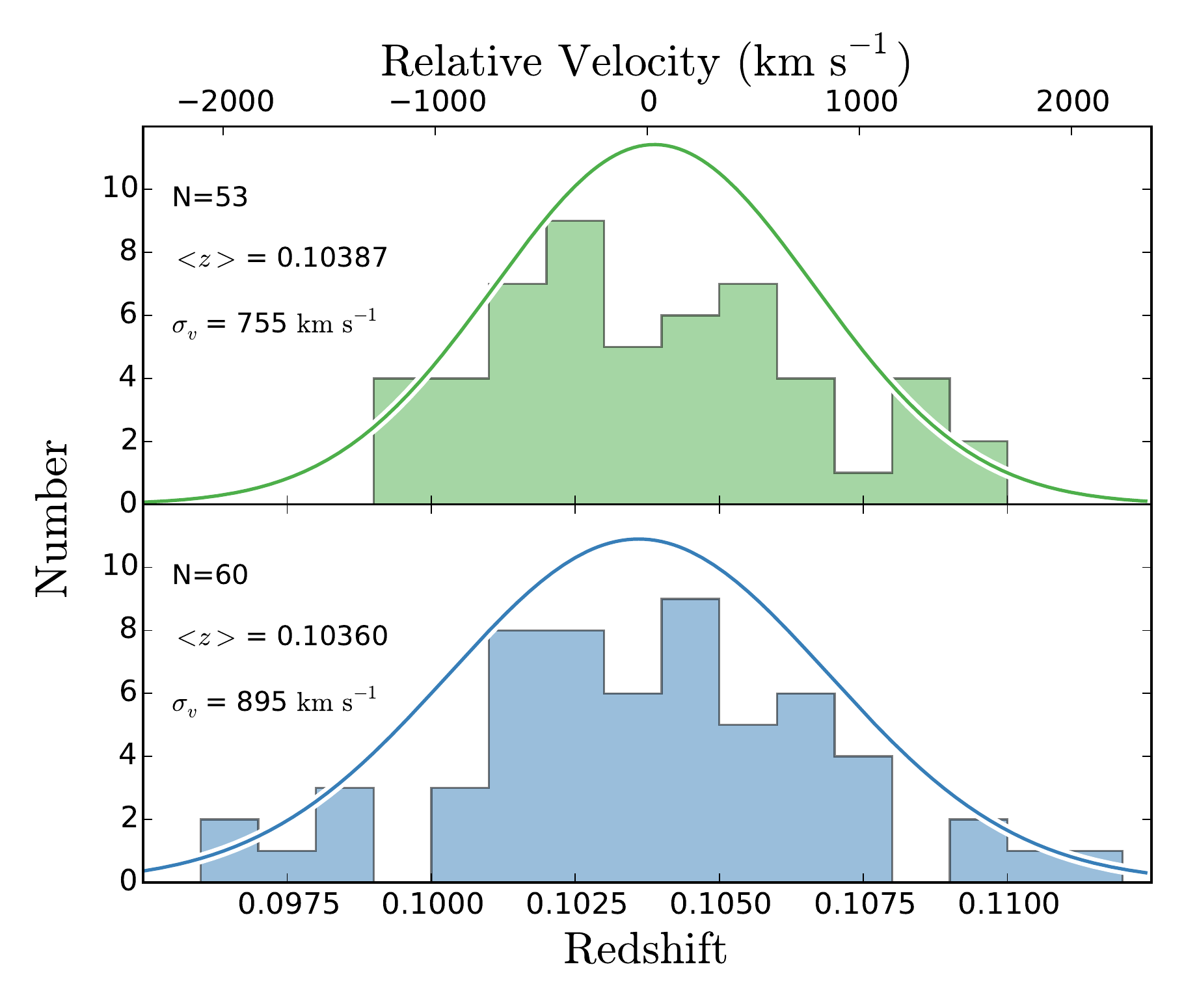}
\caption{Redshift distributions of the eastern subcluster (blue) and western subcluster (green). Redshift locations and velocity dispersions are listed in the upper-left of each panel. The east and west subcluster histograms include spectroscopic members determined by the likelihood of membership to each component of the two-component \emph{diag} model from the GMM.}
\label{fig:subclusters}
\end{figure}

Figure \ref{fig:subclusters} shows the redshift distributions of the two subclusters and accounts for this bias. The low velocity difference is only possible if the merger is occurring in the plane of the sky and/or is near apocenter. We will attempt to differentiate between the two possibilities using information from the radio relics in \S\ref{sec:mcmac}. 

The uncertainty in the bias is large despite 10$^{\text{5}}$ realizations; it comes largely from the GMM's inability to determine the center of the subclusters, which are generally known quite well given the location of dominant BCGs and the total mass distribution from the weak lensing analysis in \S\ref{sssec:massmap}. We have accounted for additional priors on the location of the subclusters with a MCMC method \citep{Golovich:2016}; however, in this cluster, the two codes gave nearly identical results, so we opted for the simpler implementation since we complete a weak lensing analysis as well. Accounting for the estimated biases, the line of sight velocity difference is $\text{92}\pm\text{164}\,\text{km}\,\text{s}^{-\text{1}}$ and the dynamical mass estimates are $\text{4.2}^{+\text{2.3}}_{-\text{1.7}}\,\times \text{10}^{\text14} \text{M}_{\odot}$ and $\text{7.6}^{+\text{4.1}}_{-\text{3.0}}\,\times \text{10}^{\text14} \text{M}_{\odot}$ for the east and west subclusters, respectively. For comparison, the entire spectroscopic catalog has a velocity dispersion of 821$^{+\text{56}}_{-\text{65}}$ km s$^{-\text{1}}$, and a corresponding dynamical mass estimate of 6.0$^{+\text{1.2}}_{-\text{1.3}}\times\text{10}^{\text{14}}\,\text{M}_{\odot}$. The total dynamical mass estimate is lower than the sum of the parts, which is a feature of splitting velocity histograms (Benson et al. in prep). These mass estimates are summarized along with all other mass estimates in Table \ref{table:masses}.

\subsection{Projected Separation}\label{subsec:2d}

\subsubsection{Galaxy Distribution}\label{sssec:densitymaps}

\begin{figure}[!htb]
\begin{center}
 \includegraphics[width=\columnwidth]{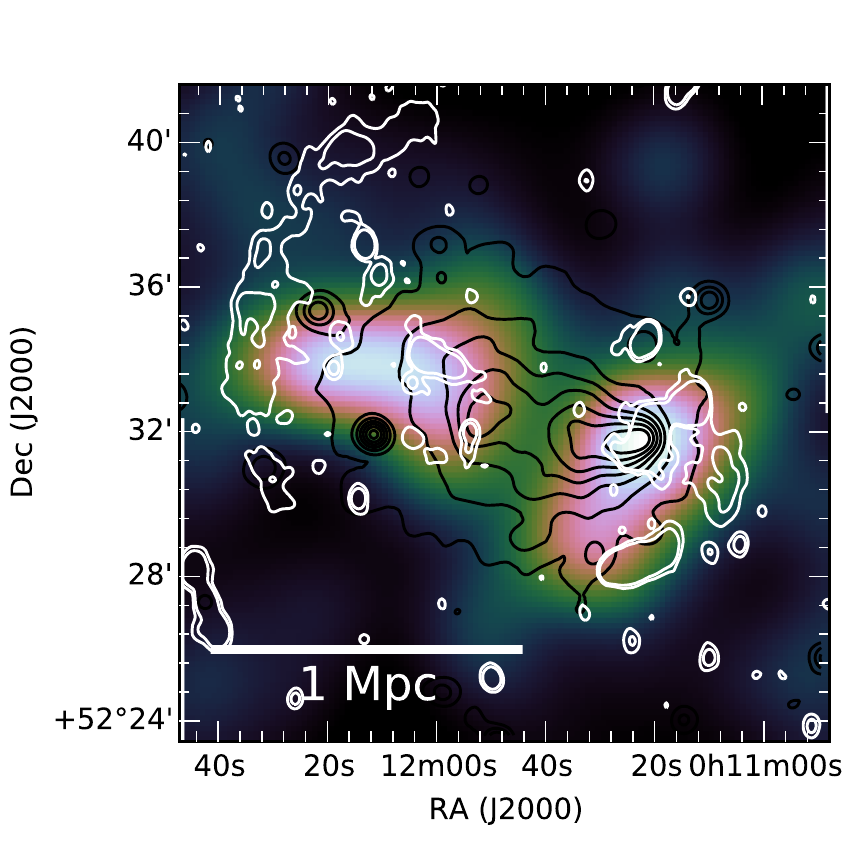} 
 \caption{Red sequence Subaru g-band luminosity distribution smoothed with a 60$\arcsec$ Gaussian kernel. The black contours are linearly scaled and represent the X-ray surface brightness distribution. White contours represent the WSRT radio data presented in \cite{vanWeeren2011}.} 
 \label{fig:density}
 \end{center}
 \end{figure}

Here we study the projected red sequence galaxy distribution. Red sequence luminosity is expected to trace the mass in the cluster \citep{Bahcall:2014}, so each galaxy is weighted by its observed r-band luminosity assuming the average distance of the cluster. We computed the optimal bandwidth for smoothing the luminosity distribution with a two dimensional Gaussian kernel using a take-one-out cross-validation scheme while maximizing the likelihood of our data under the KDE. The optimal bandwidth is $\text{96}\arcsec\times\text{52}\arcsec$. We elected to smooth the luminosity distribution with a symmetric kernel with a width of 74\arcsec, and the resulting density map is presented in Figure \ref{fig:density}. The merger axis lies mostly east-west between two dominant subclusters with the eastern subcluster slightly north of the western subcluster. This is in agreement with the radio relics and X-ray surface brightness profile (see \S \ref{sec:ICM}). The red sequence luminosity profile for the eastern subcluster is elongated east to west. This could indicate composite structure in the eastern substructure; however, the regularity of the eastern radio relic and the lack of similar structure in the X-ray surface brightness map (see Figure \ref{fig:xraymap}) suggests a bimodal merger. The projected separation between the east and west luminosity peaks is 944$^{+\text{30}}_{-\text{40}}$ kpc. To quantify the uncertainty, we perform a bootstrap analysis on the red sequence catalog. We discretize the luminosity of each galaxy in units of the least luminous galaxy, and randomly draw (with replacement) from the discretized luminosity catalog. We generate a red sequence luminosity map for each realization and measure the projected separation between the two luminosity peaks. Because the BCGs are hundreds of times brighter than the dimmest red sequence galaxy, the uncertainty is small except for the eastern subcluster, where several bright galaxies lie east to west in the vicinity of the BCG. The results of this analysis are presented in Tables \ref{table:offsets} and \ref{table:offsets1}.

\subsubsection{Mass distribution}\label{sssec:massmap}

\begin{figure*}[!htb]
\begin{center}
\includegraphics[width=\textwidth]{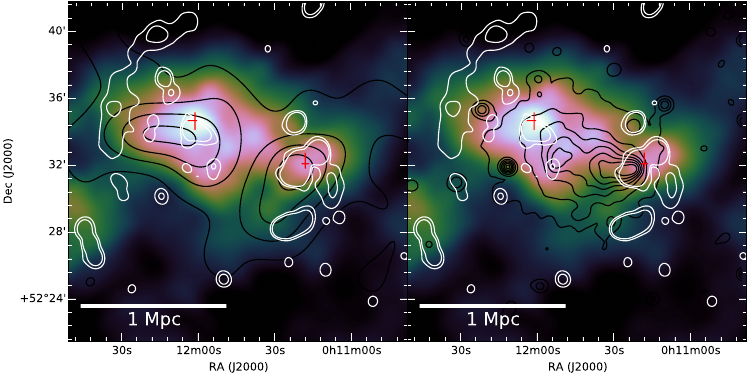}
\caption{\emph{Left:} WL mass map from a joint HST and Subaru WL analysis. Linearly spaced Subaru red sequence luminosity contours are in black. WSRT radio contours are in white. \emph{Right:} WL mass map from a joint HST and Subaru WL analysis. Linearly spaced Chandra X-ray surface brightness contours are in black. Linearly spaced WSRT radio contours are in white. The red error bars indicate the 1-$\sigma$ confidence intervals for the peak location of the two subclusters in the lensing data.}
\label{fig:massmap}
\end{center}
\end{figure*}

Here we will discuss the two dimensional mass reconstruction using both HST and Subaru data, and later in \S\ref{subsec:masses} we will discuss the mass estimation of the two subclusters. Interested readers are referred to \cite{Bartelmann:2001} and \cite{Hoekstra:2013} for more details on WL and its application to galaxy clusters; readers are referred to \cite{Jee:2014,Jee:2015} for more details regarding the method presented here. First, we perform mass reconstruction over a $\text{20}\arcmin\times\text{20}\arcmin$ region centered on the center of the Subaru field of view. The two dimensional mass reconstruction is based on the maximum entropy method of \cite{Jee07}. The method uses the entropy of the mass bins to adaptively smooth the mass map with a kernel related to the local S/N. The mass reconstruction is presented in Figure \ref{fig:massmap} and clearly shows the east-west elongation seen in the galaxies. The eastern mass peak is much broader and is generally extended along the same axis as the galaxies. Interestingly, the eastern luminosity peak extends further east than the mass peak. This is due to the BCG sitting $\sim$300 kpc east of the mass peak. Note that several bright galaxies sit to the west of the BCG as evident by the red sequence luminosity profile extending back toward the mass peak. The western mass peak is less significant, but is well aligned with the western luminosity peak. The two mass peaks lie collinearly with the two radio relics, which supports a binary, head-on merger scenario. A distinct mass peak sits southeast of the rest of the cluster, but it is not coincident with a red sequence galaxy density peak, and it is away from the spectroscopic survey. The two mass peaks are aligned with the HST fields, but we verified the location of the two mass peaks with a Subaru-only mass reconstruction. 

\begin{figure*}[!htb]
\includegraphics[width=\textwidth]{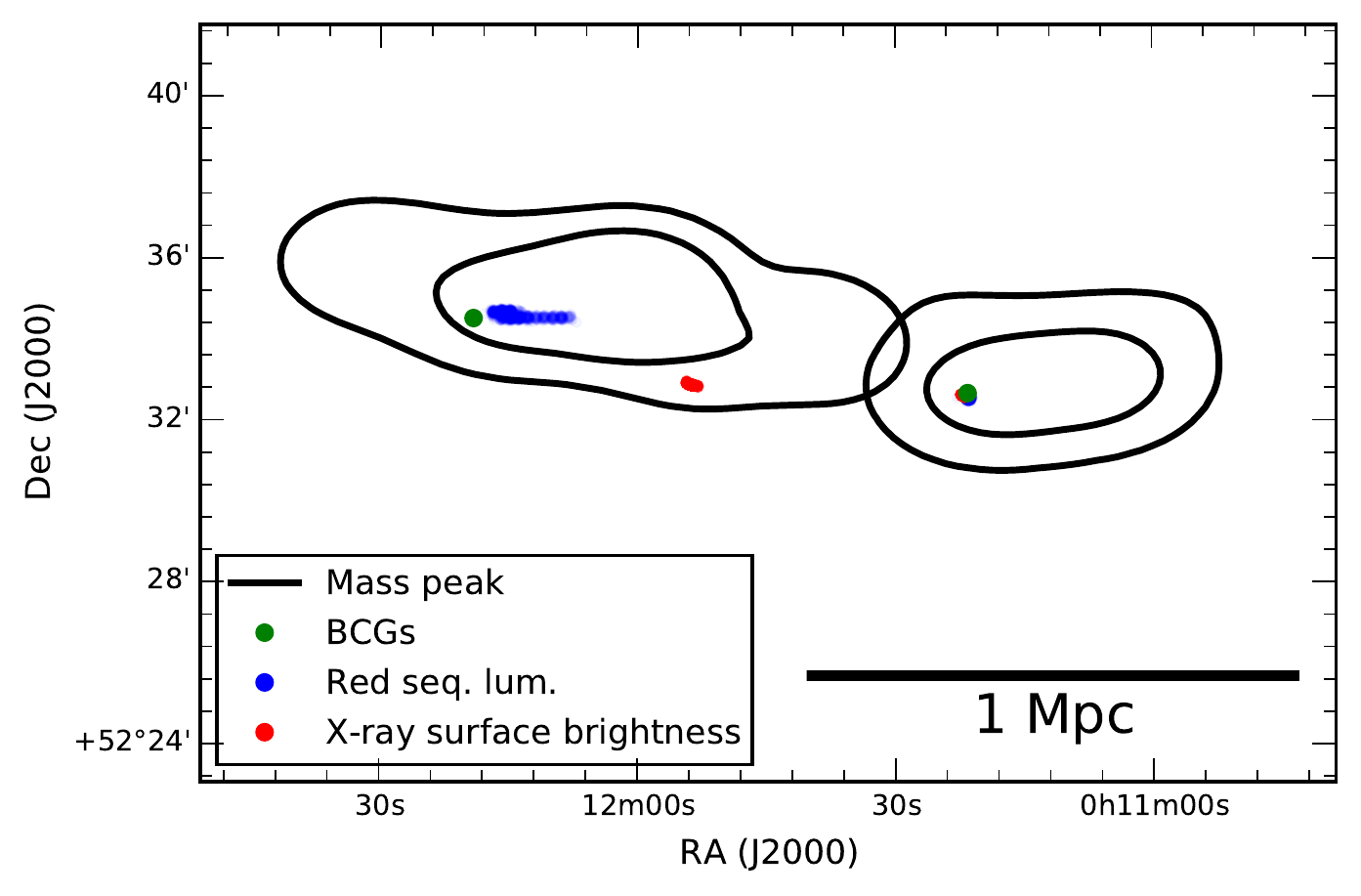}
\caption{WL peak locations from 1000 bootstrap mass maps from the source galaxy catalog. The contours show the 68$\%$ and 95$\%$ confidence regions of the peak location for the two subclusters (see \S \ref{sssec:massmap}). The bootstrapped locations for the red sequence luminosity and X-ray surface brightness data are presented as blue and red points, respectively. The two BCGs are represented with green points. There is much more variability in the locations of the eastern subcluster's components compared to the western subcluster.}
\label{fig:centroid}
\end{figure*}

To quantify the significance and uncertainty in the peak locations, we perform bootstrap analysis on the source catalog, resampling galaxies while allowing duplication and generating a mass map for each realization. We find the peak locations for the east and west subclusters in each realization and create 1- and 2-$\sigma$ confidence regions and record the projected separation for each realization. The results are shown in Figure \ref{fig:centroid} along with the red sequence luminosity, BCGs, and X-ray surface brightness peak locations in inferred from the respective bootstrap analyses. The estimated position and uncertainties of the mass peaks as well as projected separations between the east and west subclusters are summarized in Table \ref{table:offsets}. Similarly, the offsets between subcluster components are estimated with bootstrap analyses and presented in Table \ref{table:offsets1}. We will discuss these estimates further in \S\ref{subsec:offsets}. 

\begin{table*}[!htb]
\begin{center}
\caption{Subcluster component peak locations and east--west offsets}
\bgroup
\def\arraystretch{1.4}
\begin{tabular}{lccr}
\label{table:offsets}
\\
Map & East peak location (J2000) & West peak location (J2000) & Projected separation (kpc) \\
\hline  
WL massmap & 00$^{\text{h}}$12$^{\text{m}}$03.8$^{+\text{18.4}}_{-\text{11.1}}$$^{\text{s}}$, 52$^{\circ}$34$\arcmin$16.9$^{+\text{27.8}}_{-\text{32.6}}\arcsec$ &  00$^{\text{h}}$11$^{\text{m}}$12.8$^{+\text{11.5}}_{-12.5}$$^{\text{s}}$, 52$^{\circ}$32$\arcmin$11.3$^{+\text{32.1}}_{-\text{28.2}}\arcsec$  & 924$^{+\text{243}}_{-\text{206}}$  \\
Red sequence luminosity & 00$^{\text{h}}$12$^{\text{m}}$13.9$^{+\text{2.7}}_{-\text{3.8}}$$^{\text{s}}$, 52$^{\circ}$33$\arcmin$51.6$^{+\text{4.8}}_{-\text{4.2}}\arcsec$ &  00$^{\text{h}}$11$^{\text{m}}$21.6$\pm0.003$$^{\text{s}}$, 52$^{\circ}$31$\arcmin$42.1$\pm0.00002\arcsec$  & 944$^{+\text{30}}_{-\text{40}}$  \\
BCG	& 00$^{\text{h}}$12$^{\text{m}}$18.8$^{\text{s}}$, 52$^{\circ}$33$\arcmin$46.9$\arcsec$ &  00$^{\text{h}}$11$^{\text{m}}$21.7$^{\text{s}}$, 52$^{\circ}$31$\arcmin$48.5$\arcsec$  & 1020 \\
X-ray surface brightness & 00$^{\text{h}}$11$^{\text{m}}$53.7$\pm5.3$$^{\text{s}}$, 52$^{\circ}$32$\arcmin$02.2$\pm2.0\arcsec$ &  00$^{\text{h}}$11$^{\text{m}}$22.3$^{+\text{0.1}}_{-5.3}$$^{\text{s}}$, 52$^{\circ}$31$\arcmin$44.4$^{+\text{0.2}}_{-\text{0.1}}\arcsec$  & 550$\pm$4.0  \\\end{tabular}
\egroup
\end{center}
\end{table*}

\begin{table}[!htb]
\begin{center}
\caption{Offsets between subcluster components}
\bgroup
\def\arraystretch{1.4}
\begin{tabular}{lcr}
\label{table:offsets1}
Offset &	East (kpc)\footnote{Negative offsets indicate the first component listed is closer to the center of the cluster.} & West  (kpc)  \\
\hline  
DM--Luminosity &  -249$^{+\text{126}}_{-\text{141}}$  &  170$^{+\text{130}}_{-\text{131}}$\\
DM--BCG	&	-319$^{+\text{72}}_{-\text{173}}$  &  168$^{+\text{131}}_{-\text{133}}$\\
DM--ICM &  319$^{+\text{149}}_{-\text{52}}$  &  176$^{+\text{134}}_{-\text{129}}$\\
Luminosity--ICM &  420$^{+\text{20}}_{-\text{53}}$  &  12$^{+0.5}_{-\text{2.6}}$\\
BCG--ICM &  480$\pm$5.0  &  12$^{+\text{0.5}}_{-\text{2.6}}$
\end{tabular}
\egroup
\end{center}
\end{table}

\subsection{Mass Estimation}\label{subsec:masses}

In order to estimate the mass of the two subclusters, we follow the approach of \cite{Jee:2014,Jee:2015}, where the merging systems were modeled as binary systems. This is natural for ZwCl 0008, as evident from its bimodal mass and galaxy luminosity distribution and double radio relics. In \S \ref{sec:ICM} we will show further evidence for a bimodal scenario. We fit two NFW profiles simultaneously assuming the mass--concentration relation of \cite{Duffy} and fixing the centers on the two brightest galaxy luminosity peaks near the mass peak locations in Figure \ref{fig:massmap}. 

The resulting M$_{\text{200}}$ values for the east and west subclusters are 5.73$^{+\text{2.75}}_{-\text{1.81}}\times\text{10}^{\text{14}}\,\text{M}_{\odot}$ and 1.21$^{+\text{1.43}}_{-\text{0.63}}\times\text{10}^{\text{14}}\,\text{M}_{\odot}$, respectively. The masses from lensing are similar to those estimated from the velocity dispersions in scale; however, the mass ratio is reversed in that the eastern subcluster is more massive according to the lensing analysis (\S\ref{subsec:redshifts}). We expect the lensing mass to be more robust given the merging state; although, one possible explanation for the apparent discrepancy is contamination in the substructure analysis. The projected separation is small enough to ensure that some galaxies identified as members of the western subcluster are members of the eastern subcluster and visa versa, which inflates the apparent velocity dispersion. The fact that the denser gas is associated with the western subcluster lends support to it being less massive since less massive clusters have had less dynamical activity in their past. We will utilize the lensing mass estimates for the dynamical analysis (see \S\ref{sec:mcmac}).

To estimate the total mass of the system (rather than the individual subclusters), we numerically integrate the two overlapping NFW profiles in three dimensions out to $R_{\text{200}}$, taking the center to be the center of mass between the two mass peaks. We find $M_{\text{200}}=\text{8.0}^{+\text{3.6}}_{-\text{2.1}}\times\text{10}^{\text{14}}\,\text{M}_{\odot}$. We assumed the \cite{Duffy} mass--concentration scaling relations in order to generate the corresponding NFW profiles prior to integration. For the velocity dispersion mass estimates, we do not simply add the east and west masses either. Instead, we use the velocity dispersion for the entire cluster. Each of the masses estimated in this subsection are summarized and compared to other mass estimates in Table \ref{table:masses}.   

\begin{table}
\begin{center}
\caption{Subcluster and total mass estimates}
\bgroup
\def\arraystretch{1.4}
\begin{tabular}{lccr}
\label{table:masses}
Proxy & East M$_{200}$\footnote{All masses in units of 10$^{\text{14}}$ M$_{\odot}$} &  West M$_{200}$ & Total M$_{200}$ \\
\hline  
Lensing & 5.73$^{+\text{2.75}}_{-\text{1.81}}$ &  1.21$^{+\text{1.43}}_{-\text{0.63}}$ & 8.0$^{+\text{3.6}}_{-\text{2..1}}$ \\
Velocity dispersion & 4.2$^{+\text{2.3}}_{-\text{1.7}}$ & 7.6$^{+\text{3.0}}_{-\text{4.1}}$ & 6.0$^{+\text{1.2}}_{-\text{1.3}}$ \\
$L_{X}$ & - & - & 4.66$_{-\text{0.26}}^{+\text{0.31}}$ \\
$T_{X}$ & - & - & 6.5$\pm0.3$
\end{tabular}
\egroup
\end{center}
\end{table}


\section{Intra-Cluster Medium Analysis}\label{sec:ICM}

\subsection{Global X-ray Properties}\label{subsec:xray}

\begin{figure}[!htb]
\includegraphics[width=\columnwidth]{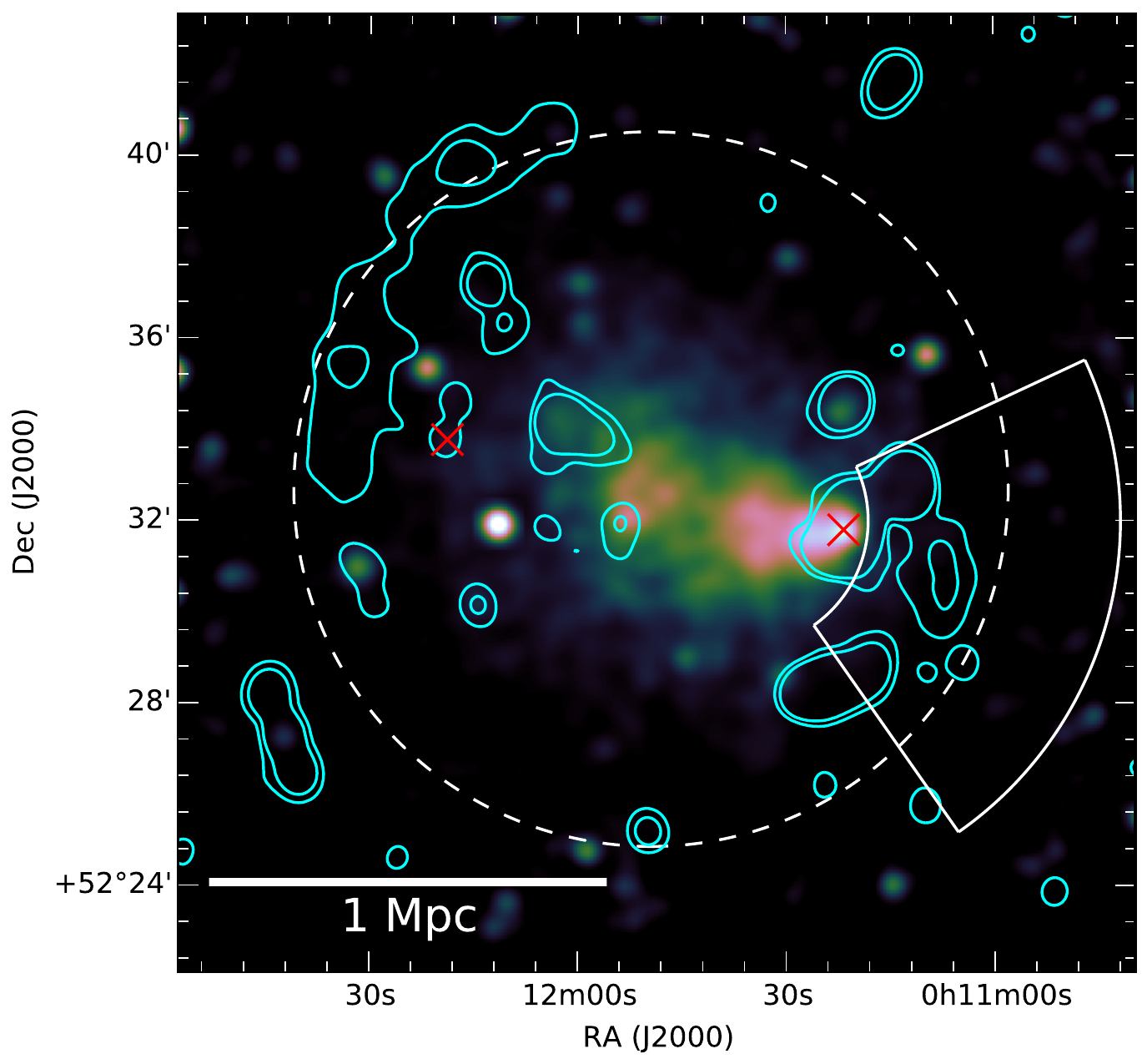}
\caption{42 ks Chandra X-ray surface brightness map (0.5--2.0 keV) for ZwCl 0008. The image was smoothed with a 3.5$\arcsec$ Gaussian kernel. The bullet and associated wake feature in the west strongly indicate an east-west merger axis. The morphology is suggestive of a cold front at the head of the bullet and a surface brightness edge associated with a shock further ahead, but the exposure resulted in insufficient counts to prove the existence of either feature. The white dashed circle represents represents the $R_{500}$ extraction region and the annular sector has an inner radius just inside of the cold front and just outside of the proposed luminosity jump including the radio relic. The two BCGs are represented with red x's. In the west the BCG and surface brightness peak are nearly coincident, and in the east there is a large offset. The cyan contours show the WSRT radio data presented in \cite{vanWeeren2011}.}
\label{fig:xraymap}
\end{figure}

The X-ray surface brightness map is presented in Figure \ref{fig:xraymap}. The morphology of the surface brightness map suggests an east-west merger between a dense remnant core (in the west in the observed state) and a more tenuous gas cloud in the east. The remnant core has a wake structure trailing behind toward the east, and there is a dense stream of gas trailing in the wake nearly  $\sim$300 kpc. The core-remnant appears to have substantially disrupted the ICM of the eastern subcluster similar to the Bullet Cluster \citep{markevitch05}. There are insufficient counts to classify the core as a cool-core remnant, so we can not make a direct comparison at this time. The core is spatially coincident with the BCG in the west, and trails the DM peak by $\sim$170 kpc. To the east, there is a distinct albeit substantially disturbed remnant core of gas, which may be what remains of the core of the east subcluster's ICM. This has been substantially offset from the BCG in the east by $\sim$320 kpc (see Table \ref{table:offsets1}).

\begin{figure}[!htb]
\includegraphics[width=\columnwidth]{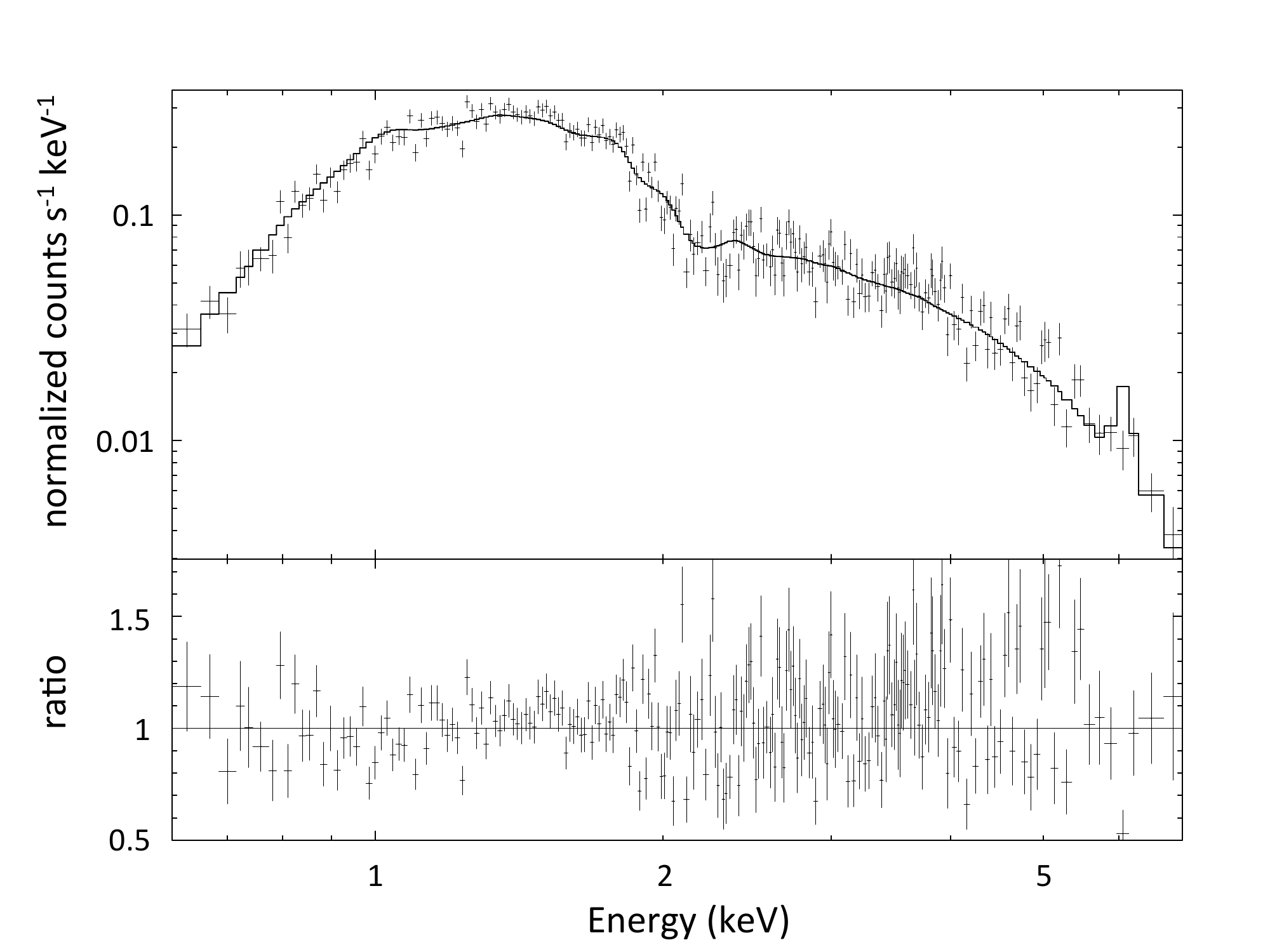}
\caption{Chandra spectrum of the $\text{R}_{\text{500}}$ region centered on the midpoint between the two BCGs of ZwCl 0008 for the full 42 ks exposure. The bottom panel shows the ratio of the data to the model.}
\label{fig:xray_spectrum}
\end{figure}

We measured the X-ray temperature and luminosity within $R_{\text{500}}$, which was estimated assuming an NFW profile with our lensing $M_{\text{200}}$ estimate and a concentration determined with the \cite{Duffy} mass--concentration relationship. This region is depicted in Figure \ref{fig:xraymap} as a dashed circle. The center of the cluster was taken to be the midpoint of the line connecting the BCGs and not the center of mass, but the temperature and luminosity derived are insensitive to the exact location of the centroid because $R_{\text{500}}$ encapsulates nearly all of the X-ray emission. We extracted all counts within $R_{\text{500}}$ and the resulting spectra were fitted with {\tt XSPEC} \citep[v12.9.0][]{Arnaud:2006}.  The $R_{\text{500}}$ spectrum and the best-fitting model are shown in Figure \ref{fig:xray_spectrum}. For the fitting we only used counts in the 0.5--7.0~keV band. The X-ray spectrum is described by an APEC model, and we fixed the metallically to a value 0.3~$Z_{\odot}$. We assume a value of 0.104 for the redshift of the cluster. The total galactic \rm{H}~\rm{I} column density was fixed to a value of $N_{\rm{H}} = 0.201\times 10^{22}$~cm$^{-2}$ \citep[the weighted average $N_{\rm{H}}$ from the Leiden/Argentine/Bonn (LAB) survey,][]{Kalberla}. We find the global X-ray temperature and luminosity to be $\text{4.9} \pm\text{0.13}$ keV and $\text{1.7}\pm\text{0.1}\times \text{10}^{\text{44}}\,\text{erg}\,\text{s}^{-\text{1}}$ in the 0.5--7.0~keV band. We also computed the X-ray luminosity in the 0.1--2.4 keV ROSAT band and found $\text{1.2} \pm \text{0.1} \times \text{10}^{\text{44}}\,\text{erg}\,\text{s}^{-\text{1}}$, which is slightly higher than the previous rough estimate base on the ROSAT count rate in \citet{vanWeeren2011}. 

These X-ray properties may be translated into mass estimates via scaling relations. For the X-ray luminosity, we use Chandra Space Telescope's \emph{PIMMS} \footnote{http://cxc.harvard.edu/toolkit/pimms.jsp} tool to translate the observed flux into a bolometric flux. We then used the \cite{Pratt:2009} scaling relation and estimate a mass of $M_{\text{500}}=\text{3.12}_{-\text{0.17}}^{+\text{0.21}}\times\text{10}^{\text{14}}\,\text{M}_{\odot}$, which translates to $M_{\text{200}}=\text{4.66}_{-\text{0.26}}^{+\text{0.31}}\times\text{10}^{\text{14}}\,\text{M}_{\odot}$ assuming an NFW profile and using the \cite{Duffy} mass--concentration scaling relations. It is still unclear if X-ray luminosity derived masses are over- or underestimates of the true mass in merging clusters. Simulations show a dependance on the viewing angle, and there is significant scatter in actual observations \citep{Takizawa:2010, Zhang:2010}. The X-ray temperature, on the other hand, is a better mass proxy for clusters. Using the \cite{Finoguenov:2001} scaling relations and assuming an NFW profile, we find the X-ray temperature scales to a mass of $M_{\text{500}}=\left(\text{4.4}\pm\text{0.2}\right)\times\text{10}^{\text{14}}\,\text{M}_{\odot}$ or $M_{\text{200}}=\left(\text{6.5}\pm\text{0.3}\right)\times\text{10}^{\text{14}}\,\text{M}_{\odot}$. The X-ray temperature scaled mass estimate is in better agreement with the WL and velocity dispersion based mass estimates. These masses are summarized and presented for comparison with the other mass estimates from this paper in Table \ref{table:masses}.

\subsection{X-ray Shocks}\label{subsec:shocks}

\begin{figure*}[!htb]
\begin{center}
\includegraphics[width=\textwidth]{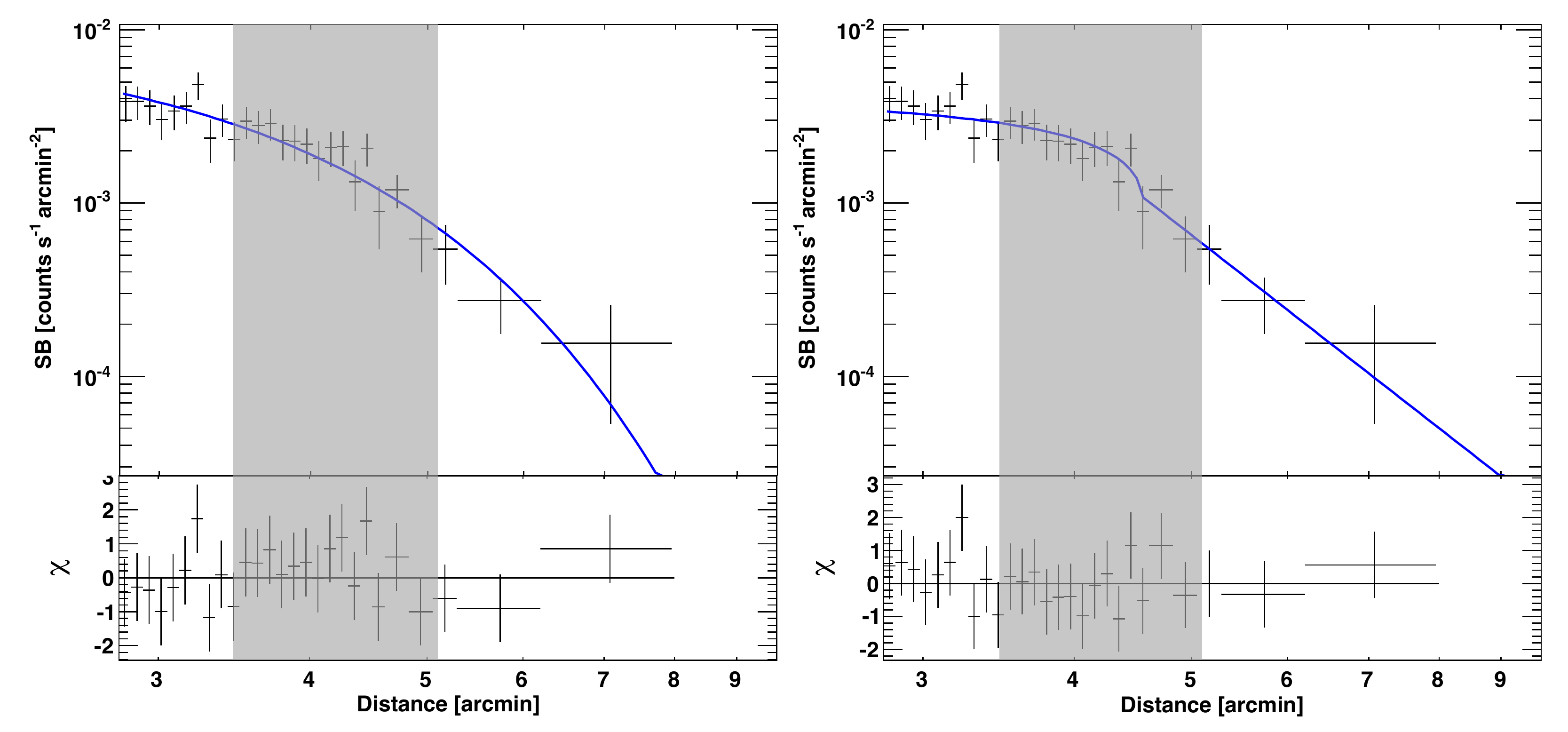}
\caption{X-ray surface brightness profile for the annular sector region of Figure \ref{fig:xraymap} with a best fit $\beta$-model (left) and shock-jump model (right). The $\beta$ model is preferred by the BIC score; although, the presence of a radio relic strongly suggests an underlying shock. The western radio relic is represented by the shaded region in each panel.}
\label{fig:shock}
\end{center}
\end{figure*}

Here we fit the X-ray surface brightness profile in the radial direction between the inner and outer radii of the sector shown in  Figure \ref{fig:xraymap} using {\tt PROFFIT} \citep{Eckert:2011}. The region is chosen to overlap a possible shock at the location of the leading edge of the west radio relic. The regions covered by the compact sources were excluded from the fit. To model the surface brightness profile, we assume the emissivity is proportional to the density squared, and this model is then integrated along the line of sight using spherical symmetry. We fit a standard $\beta$-model \citep{Cavaliere:1976} and broken power-law density model \citep[e.g.][]{Ogrean:2013c} which can be used to represent a shock in galaxy cluster outskirts. The best fitting models are shown in Figure \ref{fig:shock}. The total $\chi^{2}$ for the $\beta$ and broken power-law models are 17.0 and 13.8 with three and five model parameters, respectively. The BIC scores for the $\beta$ and broken power-law density model are 26.4 and 29.5, respectively. A $\Delta$BIC of 3.1 favors the $\beta$ model over the broken-power law model, but not strongly \citep{Kass:1995}. While we do not find evidence for the presence of a jump from these data near the western radio relic location, this is not unexpected because of the low number of X-ray counts in the radio relic region. The extra parameters of the broken power-law model carry a larger penalty than the gain offered to the $\chi^{2}$. In summary, the presence of a radio relic still strongly suggests an underlying shock, but more X-ray data are necessary to uncover it.

\subsection{Radio Analysis}\label{subsec:radio}

Large radio features (radio relics and radio halos) highlight the interaction between magnetic fields and relativistic electrons in the ICM \citep{vanweeren2010}. They are characterized by a steep radio spectrum and their placement within the cluster. Radio relics such as the two present in ZwCl 0008 appear near the outskirts of clusters and are thought to be produced by relativistic particles that have been accelerated by shocks in the ICM. Although the current X-ray data do not support the detection of shocks, a number of studies have found shocks located at the leading edge of radio relics \citep{Finoguenov:2010,Macario:2011,Shimwell:2015}. The relativistic particles in the ICM are sourced by AGN for some relics (van Weeren et al. in prep), which are then boosted by the passing shock. They emit synchrotron radiation in the presence of magnetic fields that exist in the ICM \citep[e.g.,][]{Bonafede:2010}. Radio relics are often polarized \citep{ensslin1998}, which indicates ordered underlying magnetic fields. Here we will analyze the polarization fraction of the two radio relics of ZwCl 0008. The radio observations are described in \S \ref{subsec:JVLA}.

\begin{figure*}[!htb]
\begin{center}
\includegraphics[width=\textwidth]{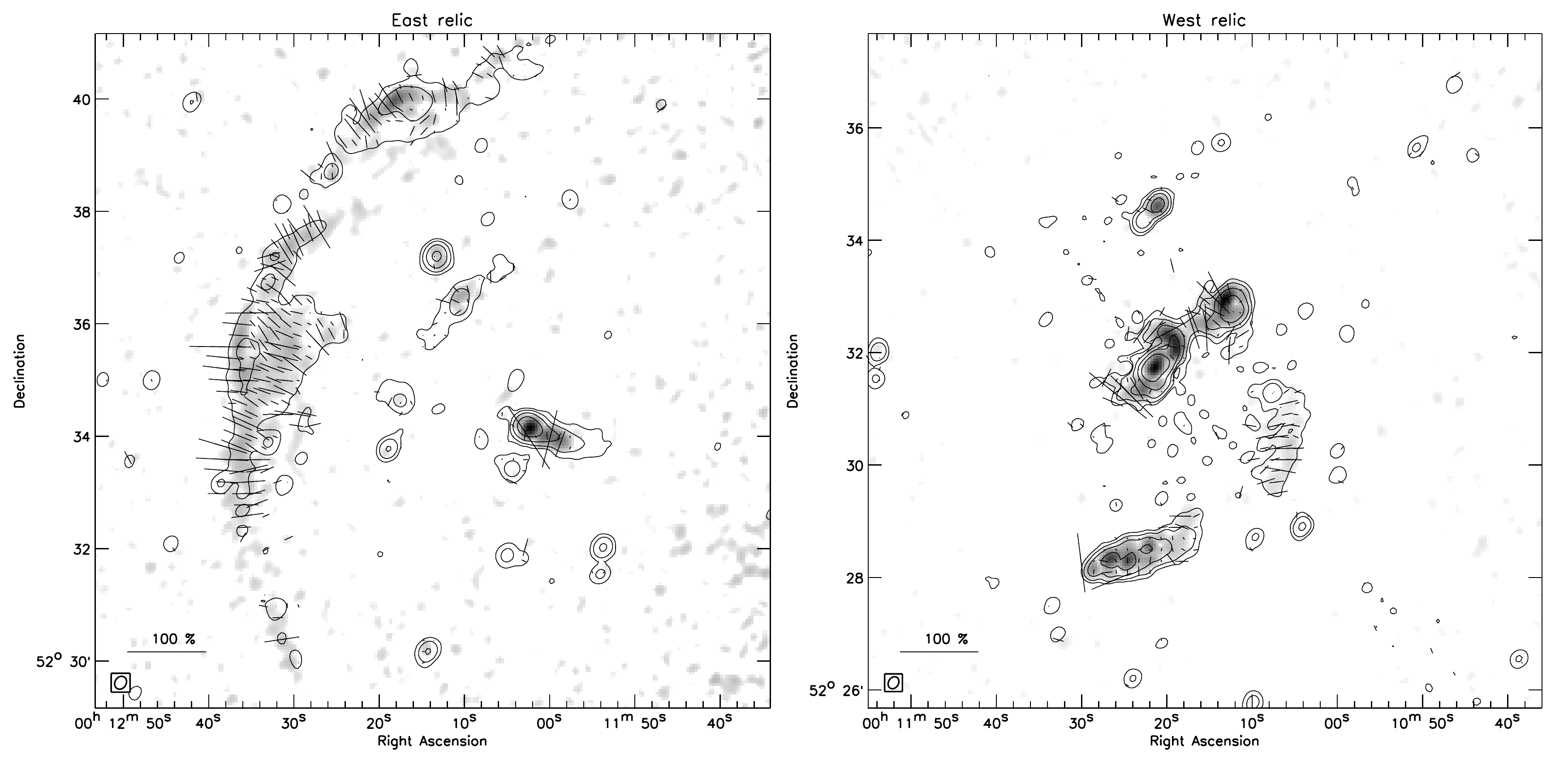}
\caption{S-band JVLA combined C and D-array image for the east (left) and west (right) radio relics of ZwCl 0008 with overlaid electric field vectors showing the polarization fraction. The eastern relic is much larger and has a higher polarization. The western relic sits just to the west of several radio loud galaxies.}
\label{fig:relic}
\end{center}
\end{figure*}

For producing the final images from the calibrated data we used WSclean \citep{Offringa:2014}. Stokes I, Q, and U images were made with the wide-band (dividing the bandwidth in eight frequency groups) clean and multi-scale algorithm. The primary beam corrected continuum images are shown in Figure \ref{fig:relic}, and are made with {\tt robust=0.5} weighting. These images have a resolution of 12$\arcsec\times$14$\arcsec$ and a r.m.s. of 8 $\mu$Jy~beam$^{-\text{1}}$. Given the galactic contribution of -30 $\text{rad}\,\text{m}^{-\text{2}}$ \citep{Taylor:2009} to the rotation measure (RM) in the direction of the cluster we can directly produce wide-band Q and U images without introducing much depolarization.  A  $\rm{RM}=30$~rad~m$^{-2}$ rotates  the polarization angle by -0.7 rad (-20$^{\circ}$ ) at 2 GHz, decreasing to -0.3 rad (-9$^{\circ}$) at 3 GHz and -0.17 rad (-5$^{\circ}$) at 4 GHz.

From the Q, U images we created images of total linear polarized intensity ($\sqrt{Q^\text{2}+U^\text{2}}$) to map the polarization fraction across the relics. A polarization vector E-field map was added to each panel of Figure \ref{fig:relic}. Polarized flux from both relics is detected and the integrated polarized fraction for the relics is 30$\%$ and 18$\%$ for the east and west relics, respectively. Locally, at a few places the polarization fraction reaches about 40$\%$, most noticeably for the eastern relic.

\section{Merger Scenario}\label{sec:scenario}

In this section, we will take stock of the evidence from the various sources to develop an understanding of the merger scenario before we study the merger dynamics in \S \ref{sec:mcmac}. 

The red sequence luminosity distribution of ZwCl 0008 displays general bimodality (see Figure \ref{fig:density}). A GMM analysis on the spectroscopic catalog (see \S\ref{subsec:subclusters}) confirms these two subclusters, and reveals them to be at very similar LOS velocities ($\Delta\text{v}=\text{92}\pm\text164\,\text{km}\,\text{s}^{-\text{1}}$), which implies that the merger is either occurring within the plane of the sky, and/or the merger is at its apocenter. A joint HST+Subaru WL analysis also confirms the two subcluster scenario (see \S\ref{sssec:massmap}) and shows the east subcluster to be $\sim$4--5 times more massive than the west subcluster (see \S\ref{subsec:masses}). The two luminosity peaks are separated by 944$^{+\text{30}}_{-\text{40}}$ kpc in projection. We analyze the mass distribution with a bootstrap analysis on the WL source catalog. We find the projected separation between the east and west peaks to be 924$^{+\text{243}}_{-\text{206}}$ kpc, which is in good agreement with the luminosity separation. The dynamics analysis in \S\ref{sec:mcmac} requires the velocities, masses, and projected separation. The velocities and masses are described above, and we use the mass separation because mass is the dynamical entity in the merger.

There is tremendous value added by the ICM analysis of \S\ref{sec:ICM} to the dynamical interpretation because it provides the direct evidence that the cluster is in a post- rather than a pre-merger stage. The most direct evidence would be from the detection of shocks in the Chandra data. In \S\ref{subsec:shocks}, we fit both a $\beta$ and shock-jump model to the X-ray radial profile in a region that overlaps the western radio relic. While the X-ray profile is modeled slightly better by a shock-jump rather than a $\beta$-model, there are insufficient X-ray counts near the relic to effectively measure the likelihood of each model with confidence. This is not to say that there is no valuable information in the X-ray data. The ``bullet'' feature in the western subcluster is clear evidence for a post-merging scenario. Furthermore, the fact that the X-ray core is so close to the west BCG is a clue to the age of the merger. The merger takes place over a couple billion years. During the outbound portion (after core passage, but before apocenter), the ICM lags the DM halo. However, after the ICM has fallen behind, there is a reacceleration phase known as the  slingshot effect, which has been observed in several cluster mergers \citep{Merten:2011,Ng:2015,Golovich:2016}. The gas-core in ZwCl 0008 is just $\sim$12 kpc away from the BCG (compared to $\sim$200 kpc in the Bullet Cluster). This suggests that the slingshot reacceleration is more advanced than the Bullet Cluster but less so than El Gordo. Perhaps there has not been enough time for the gas-core to overtake the mass peak in a slingshot scenario \citep{Hallman:2004}. We will explore this possibility further using the position of the radio relics and the output of the dynamical Monte Carlo analysis in \S\ref{sec:mcmac}. 

Despite the lack of sufficient X-ray counts to definitively detect a shock, the collinearity of the two radio relics with the two distinct subclusters provides sufficient evidence to state that ZwCl 0008 is in a post-merger stage. In \S\ref{sec:ICM}, we analyzed JVLA/C and D array data, and estimated an updated polarization fraction of 30$\%$ and 20$\%$ for the east and west relics, respectively. The observed polarization fraction depends on the viewing angle of the radio relic. \cite{ensslin1998} provides a model to place an upper limit on the angle of the merger axis with respect to the plane of the sky ($\alpha$) based on the observed polarization fraction of the radio relic. This was demonstrated in MHD simulations by \cite{Skillman:2013}, who found that it was found that face on observations resulted in observed polarization fraction of $\sim\text{10}\%$, whereas edge on observations could be as high as 60$\%$. Using the model from \cite{ensslin1998} the 30$\%$ polarization fraction of the eastern radio relic implies a merger axis within 38$^{\circ}$ of the plane of the sky. Parts of the eastern radio relic were polarized as much as 40$\%$, which implies a merger axis within 29$^{\circ}$. We will take the conservative estimate for the dynamics analysis in \S \ref{sec:mcmac}. This result is corroborated by the small LOS relative velocity of the two subclusters. The free-fall velocity at pericenter (or expected maximum relative collision velocity) using the WL masses of \S \ref{subsec:masses} is 2910 $\text{km}\,\text{s}^{-\text{1}}$. Comparing this to the low line of sight velocity difference implies that ZwCl 0008 is near apocenter and/or is merging with a substantial fraction of its merger axis parallel to the plane of the sky. 

Overall, ZwCl 0008 resembles the Bullet Cluster in that a gas-core associated with a lower mass subcluster has disrupted the more massive subcluster's ICM and stayed intact through the collision; however, the merger appears to have evolved further than the Bullet Cluster. An interesting feature is that the ICM in the east is offset far from the BCG and DM peak, whereas the west ICM peak is much more coincident with the BCG and DM, which is similar to the recent findings in Abell 1758N \citep{Monteiro:2016}. The merger axis is along an east-west axis based on the cometary shape and disrupted nature of the ICM, the bimodal mass and galaxy distributions, and the collinear radio relic with each of the subclusters.


\section{Merger Dynamics}\label{sec:mcmac}

In this section, we assemble the results of the preceding analysis to study the dynamical state of ZwCl 0008. We will use the dynamical analysis code MCMAC\footnote{\label{mcmac}{\tt https://github.com/MCTwo/MCMAC}} \citep{MCMAC} introduced and described in detail by \cite{Dawson:2012} in an analysis of the Bullet and Musketball clusters. The code has also been used on El Gordo \citep{Ng:2015}, MACS J1149.5+2223 \citep{Golovich:2016}, and Abell 3411 (van Weeren et al. in prep). The code models two NFW halos merging in otherwise empty space. We aim to estimate the timescale and velocity of the merger between the two subclusters of ZwCl 0008 as well as constrain these and geometrical properties with sampling importance resampling.

\subsection{Inputs and implementation}

\begin{figure}[!htb]
\includegraphics[width=\columnwidth]{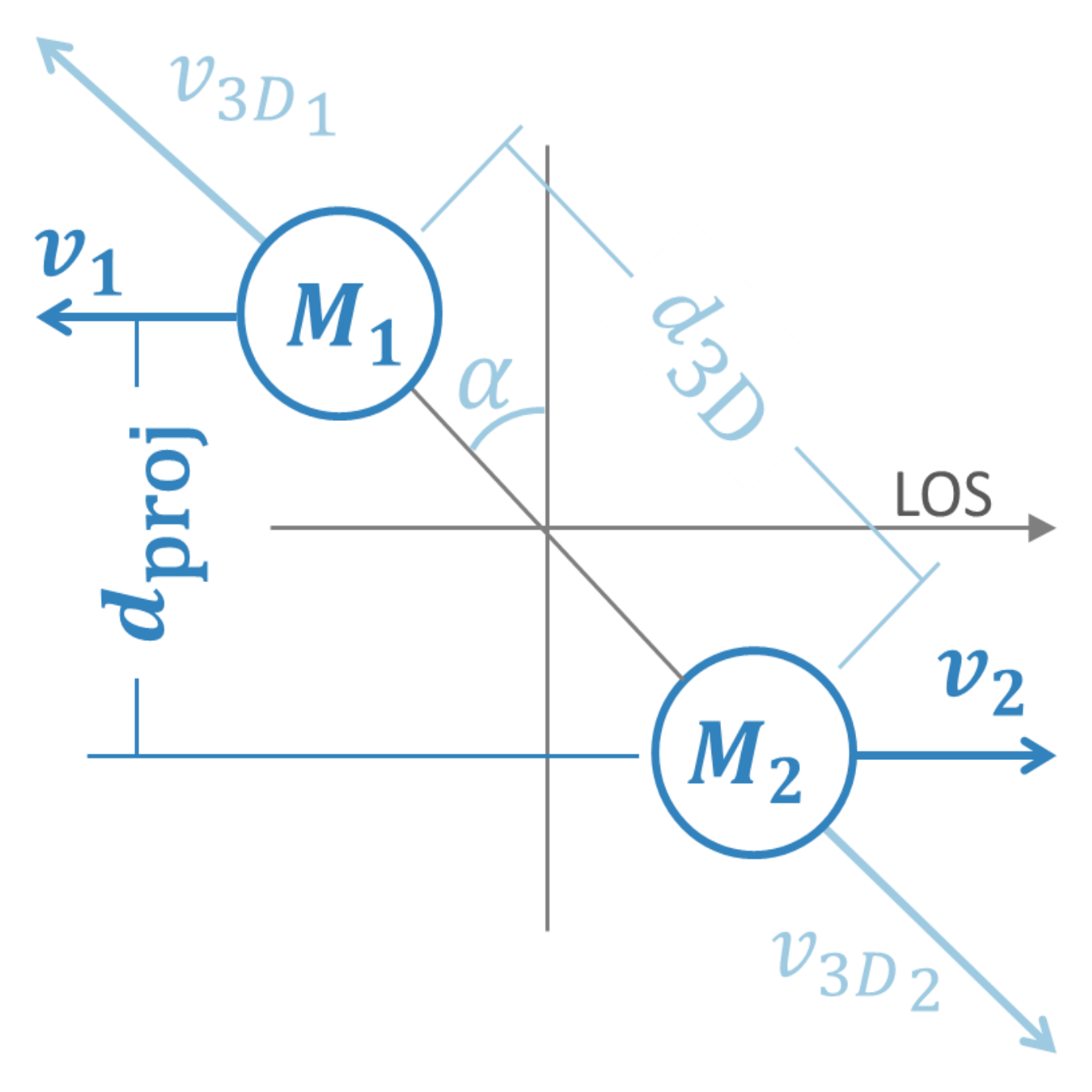}
\caption{Reproduction of Figure 1 of \cite{Dawson:2012} showing the relationships between the main input parameters for the Monte Carlo analysis. The line-of-sight is to the right, with the merger axis defined with an angle $\alpha$ measured with respect to the plane of the sky ($\alpha = 0$ corresponds to in the plane of the sky). The line-of-sight velocities ($\text{v}_{\text{1}}$ and $\text{v}_{\text{2}}$) are measured with the DEIMOS spectroscopic information, and the projected separation of the subclusters ($\text{d}_{\text{proj}}$) is measured with the Subaru optical data. $\alpha$ is constrained by the polarization of the radio relics.}
\label{fig:geometry}  
\end{figure}

MCMAC takes five inputs and their associated uncertainties generating Gaussian priors on the following parameters: the projected separation of the two subclusters in the observed state, the masses of the two subclusters, and the redshifts of the two subclusters. The parameters are defined schematically in Figure \ref{fig:geometry}. Inputs include the projected separation between the two mass peaks (see \ref{sssec:massmap}), lensing masses (see \S \ref{subsec:masses}), and subcluster redshifts (see Figure \ref{subsec:redshifts}). We run the Monte Carlo analysis until we achieve 2$\times$10$^{\text{6}}$ physical realizations. 

Posterior probability density functions (PDF) for a series of parameters (five input, three geometric, five dynamical) are output. The input parameters consist of: M$_{\text{200-E}}$, M$_{\text{200-W}}$, z$_{\text{E}}$, z$_{\text{W}}$, and d$_{\text{proj}}$. The geometry parameters consist of the randomly drawn $\alpha$, which is the angle between the plane of the sky and the merger axis ($\alpha=\text{0}$ implies parallel to the plane of the sky), calculated d$_{\text{3D}}$, and calculated d$_{\text{max}}$. The velocity/time parameters consist of: TSC$_{\text{0}}$ (time since pericenter in the outbound case), TSC$_{\text{1}}$ (time since pericenter in the returning case), T (period), v$_{\text{3D}}\left(\text{t}_{\text{col}}\right)$ and v$_{\text{3D}}\left(\text{t}_{\text{obs}}\right)$. 

\subsection{Radio Polarization Prior}

\begin{figure}[!htb]
\includegraphics[width=\columnwidth]{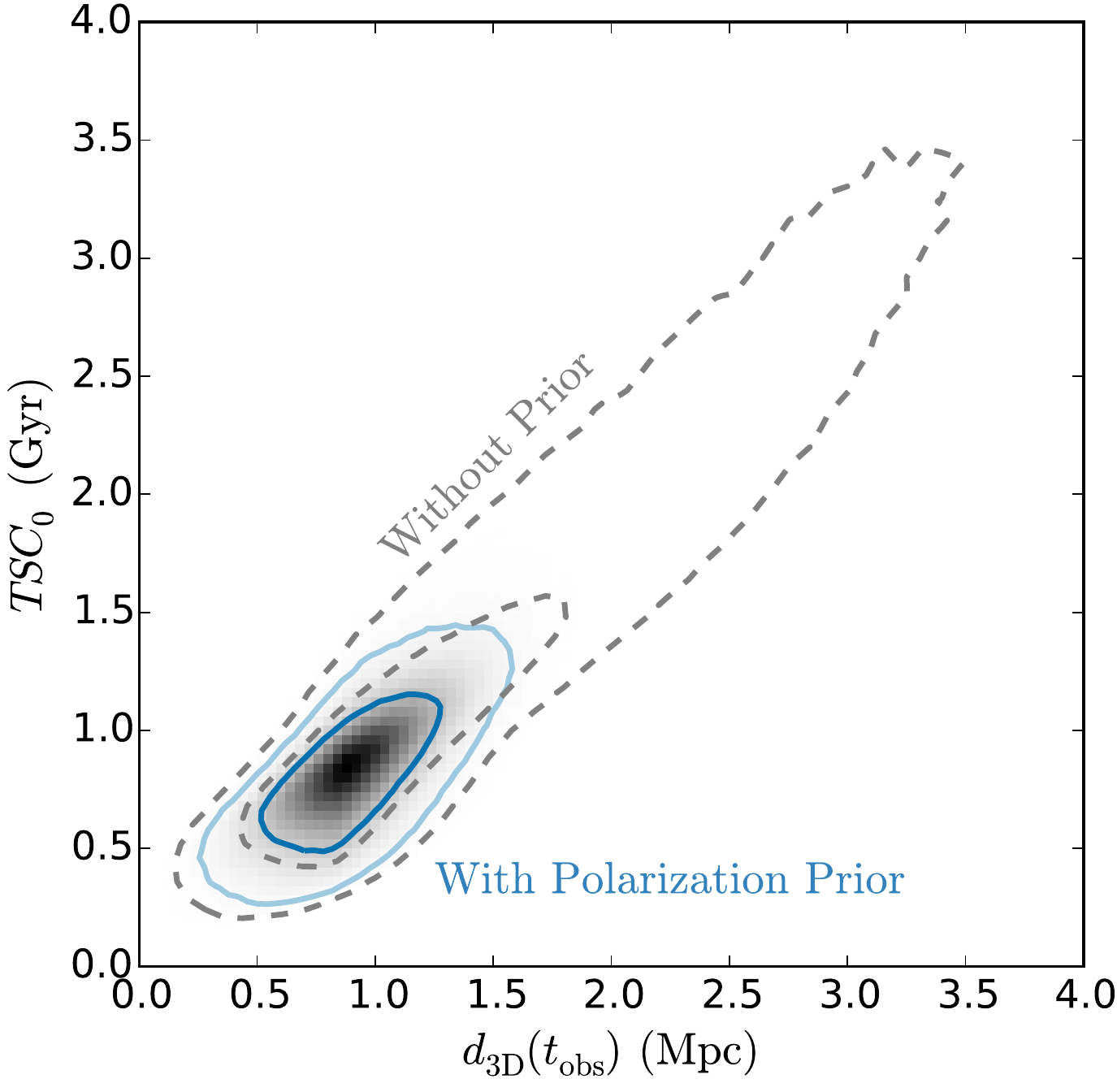}
\caption{ZwCl 0008 joint posterior PDFs for the time since collision in the outbound state versus the three dimensional separation in the observed state for the merger between the east and west subcluster. The dashed contours show the default estimates from the dynamical Monte Carlo analysis. The dark and light blue contours show the 68 $\%$ and 95$\%$ confidence limits, respectively, and show the benefit that the radio relic polarization provides.}
\label{fig:alphaprior}
\end{figure}

The largest uncertainty comes in the unconstrained value for $\alpha$ \citep{Dawson:2012}. We make use of the polarization fraction of 30$\%$ from our radio analysis in \S\ref{subsec:radio} to set an upper limit on $\alpha$ of 38$^{\circ}$ as discussed in \S \ref{sec:scenario}. This constraint greatly reduces the uncertainty in many of our estimates. Figure \ref{fig:alphaprior} shows the effect on the joint posterior PDFs for two parameters in our dynamical analysis. We summarize the output parameters and their 68$\%$ and 95$\%$ confidence limits (with the polarization prior) in Table ~\ref{table:mcmac}. 

\begin{table}[!htb]
\begin{center}
\caption{ZwCl 0008 merger parameter estimates including the polarization prior} 
\bgroup
\def\arraystretch{1.4}
\begin{tabular}{llll}
Parameter\footnote{\label{parameters}$M_{\text{200}}$ mass; $z$ redshift; $d_{\text{proj}}$ projected subcluster separation; $\alpha$ angle between the merger axis and the plane of the sky; $d_{\text{3D}}$ 4-D subcluster separation; $d_{\text{max}}$ 3-D maximum subcluster separation after core passage; $v_{\text{3D}}\,\left(t_{\text{obs}}\right)$ subcluster relative velocity in the observed state; $v_{\text{3D}}\,\left(t_{\text{col}}\right)$ subcluster relative velocity at core passage; $TSC_{\text{0}}$ time since collision for the outbound scenario;  $TSC_{\text{1}}$ time since collision for the return scenario; $T$ time between collisions. See \cite{Dawson:2012} for more details on these quantities.}	& Location\footnote{\label{biweight}Biweight-statistic location \citep{Beers1990}}	& 68$\%$ LCL--UCL\footnote{\label{bcpcl} Bias-corrected lower and upper confidence limits \citep{Beers1990}}	&   Units\\
\hline 
$M_{\text{200}_{\text{E}}}$		& 5.9			& 3.3--8.6			& $\text{10}^{\text{14}}$\,\text{M}$_{\sun}$\\	
$M_{\text{200}_{\text{W}}}$		& 1.7			& 0.69--3.0		& $\text{10}^{\text{14}}$\,\text{M}$_{\sun}$\\	
$z_{\text{E}}$					& 0.10363		& 0.10321--0.10404	& \\	
$z_{\text{W}}$					& 0.10385		& 0.10349--0.10421	& \\	
$d_{\rm proj}$					& 0.83		& 0.64--1.1		& \text{Mpc}\\ 	
$\alpha$						& 17			& 6.6--31			& degrees\\ 	
$d_{\rm 3D}$					& 0.89		& 0.67--1.1		& \text{Mpc}\\ 		
$d_{\rm max}$					& 1.0			& 0.74--1.4		& \text{Mpc}\\ 		
$v_{\rm 3D}(t_{\rm obs})$			& 100		& 33--240			& \text{km}\,\text{s}$^{-\text{1}}$\\ 	
$v_{\rm 3D}(t_{\rm col})$			& 1800		& 1500--2200		& \text{km}\,\text{s}$^{-\text{1}}$\\ 	
$TSC_{0}$					& 0.76		& 0.49-1.0			& \text{Gyr}\\ 	
$TSC_{1}$ 					& 1.3			& 0.95--2.2		& \text{Gyr}\\ 	
$T$							& 2.1			& 1.7--2.9			& \text{Gyr}\\ 
\end{tabular}
\label{table:mcmac}
\egroup
\end{center}
\end{table}

\subsection{Radio Relic Locations}\label{subsec:relics}

The dynamical analysis utilized in this section does not take into account dynamical friction; thus, the equations of motion are time-reversible and degenerate between a still outbound and returning scenario for a given random realization. We have mentioned that the X-ray gas-core in the west is much more intact and nearly coincident with the BCG in the western subcluster, which could suggest a merger near turnaround if this is interpreted as an intermediate stage of the slingshot effect \citep{Hallman:2004}. Here we will make use of the radio relics to attempt to break this degeneracy and further constrain the dynamical parameter estimates.

For each Monte Carlo realization, the center of mass is computed and fixed to the inferred merger axis defined by the line connecting the two mass peaks in the mass map in Figure \ref{fig:massmap}. The projected distance from the center of mass in each realization to the leading and trailing edge of the relics is measured. The observed position of the relics is compared to a calculated relic position estimated following the formalism developed in \cite{Ng:2015}. The output realizations for the MCMAC analysis are used to predict the expected location of the radio relics. ICM shocks from cluster mergers are thought to be generated near the center of mass during core passage and propagate outward with a speed similar to the merger velocity of the two subclusters \citep{springel2007,Paul:2011, vanWeeren:2011}. \cite{springel2007} showed for the Bullet Cluster, the time-averaged shock propagation speed is 90$\%$ of the collision speed. For ZwCl 0008, the merger age is older. We extrapolated the results of \cite{springel2007} and estimate the time-averaged shock propagation speed to be 70--90$\%$ of the collision speed. We represent decrement in shock propagation speed with respect to the collision speed as a factor $\beta$, which we randomly draw from $\mathcal{U}\left(0.7,0.9\right)$. We calculate the expected projected shock propagation distance (in the center of mass frame) for the two relics for each of the two age estimates from our dynamical analysis:\begin{equation}\label{eq:outbound}
\text{s}_{\text{i}} = \frac{\text{M}_{\text{j}}}{\text{M}_{\text{i}}+\text{M}_{\text{j}}}\,\beta\,\text{v}_{\text{3D}}\left(\text{t}_{\text{obs}}\right)\, \text{TSM}\,\cos{\alpha}
\end{equation}
where $i$ and $j$ refer to the two subclusters, $\beta$ is drawn from $\mathcal{U}\left(\text{0.6},\text{0.9}\right)$, $\text{v}_{\text{3D}}$ is the three dimensional collision velocity of the two subclusters, $\text{TSM}$ is the time since merger pericenter in either the outbound or return case, and $\alpha$ is the angle between the merger axis and the plane of the sky.

\begin{figure}[!htb]
\includegraphics[width=\columnwidth]{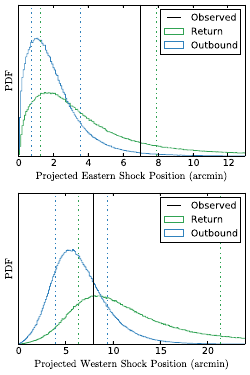}
\caption{\emph{Top:} Predicted and observed shock position for the eastern subcluster of ZwCl 0008 in the center of mass frame. The posteriors for the predicted positions are estimated from the MCMAC outputs. The observed position is measured from the center of mass to the leading edge of the eastern radio relic. The dot-dash lines show the 68$\%$ confidence region centered on the median for each model. \emph{Bottom:} Same analysis for the western subcluster. }
\label{fig:relicposition}
\end{figure}

These values are calculated for each realization output by MCMAC and are normalized and plotted in Figure \ref{fig:relicposition}. The black lines indicate the observed position of the leading edge of the radio relic (where the shock is assumed to be located) with respect to the center of mass, and the blue and green curves represent the normalized posterior PDFs for the expected shock position using the MCMAC estimates. 

The eastern shock is far from the center of mass and puts slight tension on the outbound scenario. This could be due to a number of factors, but possibilities include issues with mass estimation and location of the two DM haloes related to the low lensing signal of our photometry. Also, perhaps the propagation of the two shocks is more complicated than our one parameter ($\beta$) model can describe. Meanwhile, the west relic's observed position is nearly equally described by the outbound or return scenarios.

To check the consistency to our observations of this simple one-parameter model of radio relic propagation, we kept only Monte Carlo realizations that satisfy the shock locations. The relic locations require a mass ratio of 2:1, which is not ruled out by our lensing results with 1-$\sigma$ confidence. 

\section{Discussion}\label{sec:discussion}

In this section, we discuss our findings in relation to previous studies of ZwCl 0008 and in comparison to other systems that we have studied. We expand upon our findings regarding offsets between subcluster components and the merger phase, which is important to quantify in order to understand the offsets in regards to alternate models of DM. We also consider this merger's place among other double radio relic clusters. 

\subsection{Comparison to Previous Studies of ZwCl 0008}

Three previous studies of ZwCl 0008 are in the literature. Two of these are radio analyses. \citet{vanWeeren2011} first presented a merging picture based on the discovery of two radio relics from GMRT and WSRT data and bimodal photometric galaxy distribution based on INT imaging. The radio relics are revealed to be polarized at $\sim$25$\%$ with spectral steepening towards the cluster center. Meanwhile, \citet{Kierdorf:2016} recently studied the east radio relic of ZwCl 0008 at high radio frequencies. Our JVLA data reveal slightly higher polarization values, which we attribute to depolarization in the lower frequency WSRT observations. Meanwhile, \citet{Kierdorf:2016} mentions that deeper observations at the high radio frequencies that they analyzed are necessary to fully map the polarization structure.\citet{vanWeeren2011} found the relics to indicate Mach numbers of ($\mathcal{M}$) of $\text{2.2}_{-\text{0.1}}^{+\text{0.2}}$ and $\text{2.4}_{-\text{0.2}}^{+\text{0.4}}$ for the east and west relics, respectively. If we assume a uniform global temperature, the shock propagation speed in a $\sim$5 keV gas is $\sim$1150 km s$^{-1}$, which corresponds to shock propagation speeds of $\sim$2500 km s$^{-1}$ and 2750 km s$^{-1}$ for the east and west relic, respectively. These propagation speeds represent the speed of the (presumed) shock at the leading edge of the relics \emph{relative to the frame of the medium} and not the merger speed, which should be lower since the medium is likely flowing inward against the shock. The merger speed was estimated in \S\ref{sec:mcmac} to be $\text{1800}^{+\text{400}}_{-\text{300}}\,\text{km}\,\text{s}^{-\text{1}}$.

\citet{Kang:2012} studied ZwCl 0008 via diffusive shock acceleration simulations and found that a projection angle between 25 and 30$^{\circ}$ to best model the spectral index and radio flux. We estimated the angle to be $\text{17}^{+\text{14}}_{-\text{10}}$$^{\circ}$ from the plane of the sky. Furthermore, using the radio relic polarization, we set an upper limit on the viewing angle of 38$^{\circ}$. Our results are in good agreement with \citet{Kang:2012}. 

\subsection{Comparisons to Other MCMAC Studies}

\begin{figure}[!htb]
\includegraphics[width=\columnwidth]{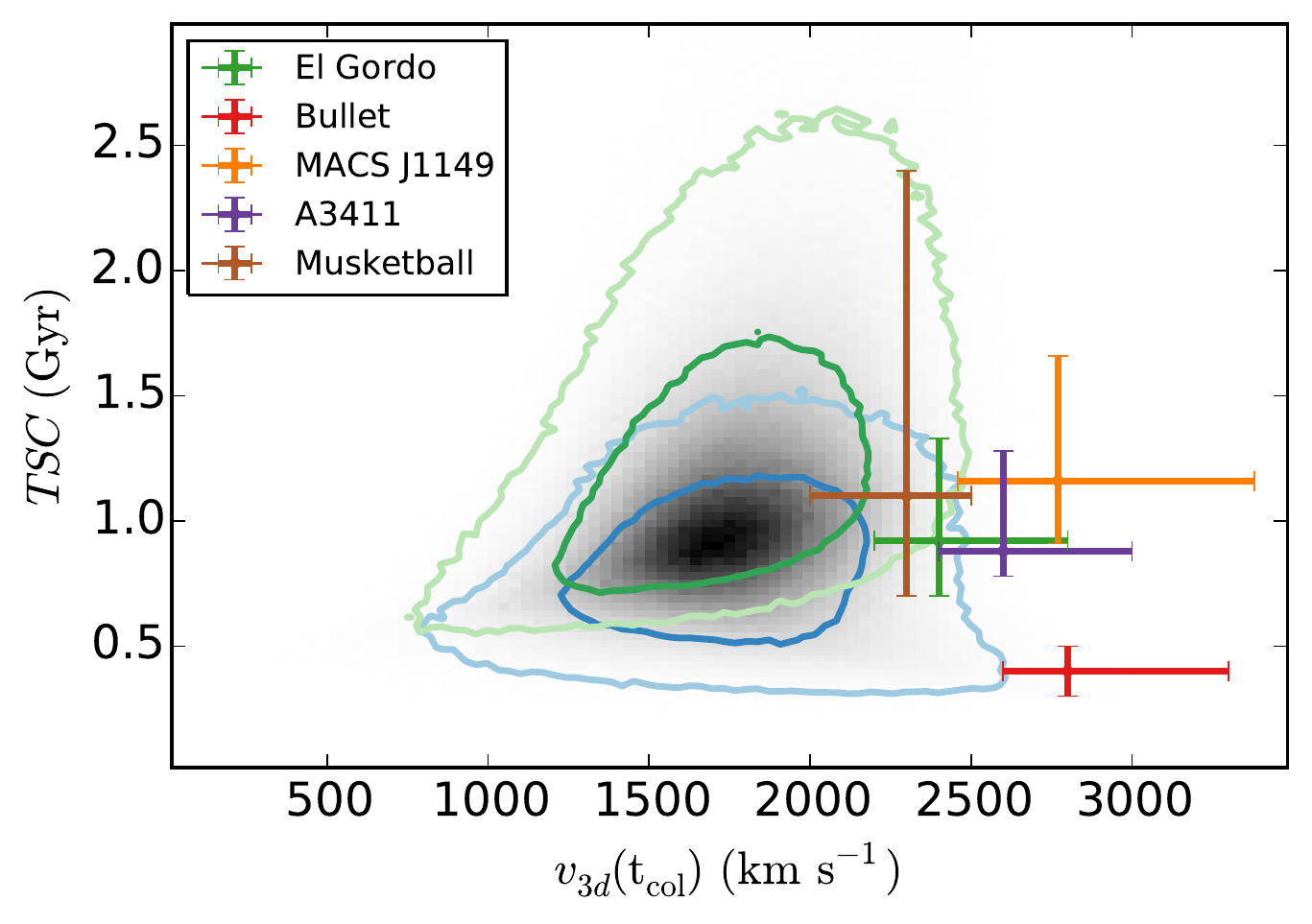}
\caption{The joint posterior PDFs for three dimensional merger velocity and time since collision in the outbound and return cases. Dark and light blue (green) contours represent 68\% and 95\% confidence regions for the outbound (return) scenario, respectively. The values for the same parameters for the Bullet Cluster, MACS J1149, Abell 3411, and the Musketball clusters  (green, red, orange, purple, and brown, respectively), which were analyzed similarly \citep{Dawson:2012,Golovich:2016, Ng:2015} are overlaid for comparison. The gray shading corresponds to the combined likelihood for the outbound and return scenarios, which we are unable to select between.}
\label{fig:TSM0compare}
\end{figure}

We have studied six mergers with the dynamical analysis technique used in \S\ref{sec:mcmac}. Each of these clusters are major mergers, and ZwCl 0008 less massive than each except the Musketball Cluster. ZwCl 0008 is $\sim\text{50}\%$ of the mass of the other clusters. Figure \ref{fig:TSM0compare} shows the relationship between joint posterior PDFs for the time since collision and the three dimensional collision speed. Also plotted are the 68$\%$ confidence regions for MACS J1149 \citep{Golovich:2016}, the Bullet and Musketball clusters \citep{Dawson:2012}, El Gordo \citep{Ng:2015}, and Abell 3411 (van Weeren et al. in prep). 

ZwCl 0008 is most similar to the Bullet Cluster morphologically; however, it occupies a slower and older portion of dynamical phase space. This makes sense since higher mass inflates the merger speed due to the larger gravitational attraction between the subclusters before pericenter. The Musketball cluster has a higher merger speed despite its lower mass, but we note that the dynamical analysis is highly uncertain due to its low mass and lack of radio data to further constrain the results. 

\subsection{Double Radio Relic Clusters}

The presence of two radio relics on opposite sides of ZwCl 0008 provides substantial evidence that the merger is largely bimodal and head on \citep{vanWeeren:2011b}. This feature validates the zero impact parameter assumption made in our dynamics analysis. It also ensures maximal interaction between the passing DM and ICM and thus substantial disruption and formation of offsets. 

\citet{deGasperin:2014} provides a list of 15 such systems (including ZwCl 0008) and studied them for correlations between mass, radio luminosity, relic position, morphology, and redshift. For the mass relations, \citet{deGasperin:2014} utilized the M$_{500}$ estimates derived from the SZ effect by the Planck Collaboration \citep{PlanckMass} for all but three clusters, which are not present in the Planck catalog. For those three clusters, they used X-ray luminosity measurements and scaling relations from \citet{Pratt:2009}. The presence of radio relics strongly implies associated shocks traveling in the ICM. These shocks transfer heat more efficiently to protons, which then come to equilibrium with electrons in the post-shock region through coulomb coupling over the course of $\text{10}^{\text{7--9}}$ Yrs \citep{Takizawa:1999}. Thus, mass proxies based on empirical scaling relations of ICM observables are inaccurate for merging clusters as the gas is neither in thermal equilibrium with itself or in gravitational equilibrium with the DM halo. Furthermore, since the ICM reproaches thermal equilibrium, the accuracy of these mass proxies varies. Therefore, gravitational lensing is preferable in these cases since the lensing signal is only dependent on mass and distances. 

We use the \citet{Duffy} mass--concentration scaling relations to convert our $M_{200}$ mass estimate from our lensing analysis to $M_{500}$. For ZwCl 0008, we find $M_{500}\sim\text{5.5}\times\text{10}^{\text{14}}\,\text{M}_{\odot}$ compared to $M_{500}=\text{3.3}^{+\text{0.46}}_{-\text{0.50}}\times\text{10}^{\text{14}}\,\text{M}_{\odot}$ from the Planck catalog. This verifies the value of updating the rest of the mass estimates where possible. We also note that since \citet{deGasperin:2014} was published, three additional double radio relic systems have been identified. PSZ1 G108.18-11.53 was recently discovered to have two powerful radio relics \citep{deGasperin:2015}; additionally, with the help of low-frequency radio observations with LOFAR, both Abell 2034 \citep{Shimwell:2016} and the Toothbrush Cluster \citep{vanWeeren:2016} have been shown to contain multiple radio relics. Most recently, \citet{Riseley:2017} show MACS J0025.4-1222 to contain two radio relics, which brings the total to 19 known systems hosting multiple radio relics.

\subsection{Interpreting the Offsets}\label{subsec:offsets}

In Table \ref{table:offsets} we presented peak locations and projected separations between the east and west subcluster components of four location proxies (lensing peak, red sequence luminosity peak, BCG location, and X-ray peak). For the east--west offsets, the DM and galaxy proxies give consistent estimates of $\sim$900 --1000 kpc between the east and west subcluster centers. These offsets should agree assuming $\Lambda$CDM. Meanwhile, the ICM interacts during the merger much more strongly, so we expect significant departure in the distance between the remnant gas peaks as compared to the DM and galaxies. This discrepancy is what motivated the term \emph{dissociative cluster} to describe this phase of cluster formation. 

Note that the findings in merging clusters are in sharp contrast with relaxed clusters. Ng et al. (in prep) analyzed an ensemble of halos in the Illustris simulation \citep{Illustris} and found that the BCG is the most accurate indicator of cluster peak locations among galaxy location proxies when compared to peak of the gravitational potential. Ng et al. (in prep) also examined galaxy luminosity density and galaxy number density and found that the luminosity density is nearly as accurate as the BCG location, but the number density distribution is a poor proxy for locating the galaxy distribution in relation to the gravitational potential. We examined offsets between the BCGs, red sequence luminosity distribution, lensing mass distribution, and X-ray surface brightness distributions. These are presented in Table \ref{table:offsets1} for each subcluster. 

As expected, we find large offsets between the lensing mass and ICM peak locations in each subcluster. The total mass leads the gas in each subcluster substantially. The galaxy (luminosity and BCG) proxies are notably discrepant from the lensing position, which is unexpected in $\Lambda$CDM; however, due to the low lensing signal in our imaging data, these offsets are statistically insignificant. In the east, the total mass trails the red sequence luminosity peak by 249$^{+\text{126}}_{-\text{141}}$ kpc and the BCG by 319$^{+\text{72}}_{-\text{173}}$ kpc. Meanwhile, in the west, the total mass \emph{leads} both the luminosity distribution peak and the BCG by 170$^{+\text{130}}_{-\text{131}}$ kpc and 168$^{+\text{131}}_{-\text{133}}$ kpc, respectively. It would unusual to see offsets with mixed signs such as this where the DM trails the BCG in one subcluster but leads the BCG in the other subcluster. Deeper imaging will be needed to probe these offsets more thoroughly. In $\Lambda$CDM, it is expected for the galaxies and DM to coincide. \citet{Kim:2016} found the DM to lead the galaxies in their self-interacting dark matter (SIDM) simulations of during later phases of bimodal mergers (after apocenter as the subclusters approach for a second core passage), but these offsets were an order of magnitude smaller than the measured offsets in ZwCl 0008. Also, we note that these simulations were between equal mass subclusters and do not include hydrodynamics; however, they are the most detailed examination yet of the complete time evolution of the DM--galaxy offsets.

The offset between the BCG and luminosity peak in the east is also interesting. We note that it is biased by our choice of smoothing kernel for the red sequence density map, which was selected to maximize the likelihood of our data under the KDE via a take-one-out cross-validation scheme. The optimal kernel was was 96$\arcsec$$\times$52$\arcsec$ along the RA and DEC axes. We elected to smooth the luminosity distribution with a circular 74$\arcsec$ kernel; however, if we had instead chosen a 52$\arcsec$ kernel, there would be two sub-peaks in the luminosity distribution in the east. No other evidence points to composite substructure in the east, so we chose against a 52$\arcsec$ kernel. With a 96$\arcsec$ smoothing kernel, the east--west alignment of the subclusters began to disappear as the luminosity peaks become blended. 

Also notable is that the ICM and galaxy distributions are in close agreement in the west subcluster but not in the east subcluster. This is reminiscent of Abell 1758N \citep{Monteiro:2016}.

Previous studies have utilized offsets between the galaxies, DM, and gas to derive upper limits on $\sigma_{\text{SIDM}}$ \citep[see e.g.,][]{randall2008,Merten:2011}. ZwCl 0008 is a suitable system for such a calculation; however, the relatively low mass and redshift lowers the lensing signal compared to other clusters that we have examined with similar techniques \citep[e.g., the Sausage and Toothbrush clusters:][respectively]{Jee:2015,Jee:2016}. The lower lensing signal limits our ability to pin down these offsets for a reliable independent estimate. It would be beneficial to extend the HST observations and obtain deeper imaging at all wavelengths to more precisely model the ICM, DM, and galaxy distributions as necessary to investigate these offsets individually for an SIDM signal. The Merging Cluster Collaboration is analyzing a sample of 29 merging clusters to study as an ensemble, which will allow for a more robust investigation of these offsets in merging clusters. We will include these two subclusters in our future ensemble analysis.

\subsection{Merger Phase}

Characterization of the phase of the merger is an important task if clusters are to be used to constrain SIDM \citep{Kim:2016}. In SIDM models, offsets between the galaxy and DM distributions of subclusters grow and shrink over the life of the merger, but in the time between the first pericenter crossing and apocenter the DM is expected to fall behind the galaxy distribution before the subclusters slow and gravity takes over. Even in $\Lambda$CDM, the time evolution of offsets between the DM and ICM are time dependent, so no matter the model for DM, it is important to understand the phase of the merger. Here we discuss our ability to constrain the phase of ZwCl 0008. 

First, we used the Monte Carlo samples from our dynamical analysis in conjunction with the location of the radio relics to select between outbound and return models (see \S\ref{subsec:relics}). This analysis gave unclear results for the phase of the merger. The west radio relic location is equally described by outbound or returning scenarios, but the east radio relic's location is marginally better described by a return scenario. Due to this discrepancy, we seek a second method to constrain the phase of the merger.

Using the MCMAC parameters for the time since collision and the period, we define a dimensionless phase: $\tau = \frac{TSC}{T}$. The Bullet Cluster is clearly in the outbound scenario based on the location of the shocks and has $\tau=\text{0.27}$. This is also clear from the X-ray surface brightness morphology and the location of the cool-core, which is $\sim$80 kpc behind the mass peak. Assuming the outbound state, ZwCl 0008 has $\tau = \text{0.38}$, but the shock positions are inconclusive for model selection. However, ZwCl 0008 contains a gas-core remnant that has a very similar morphology as the Bullet Cluster. It appears to be traveling to the west, yet it is nearly coincident with the BCG (trails by just 12 kpc). The offset with the mass peak is much larger (176$^{+\text{134}}_{-\text{129}}$ kpc). The large uncertainty is due to our low lensing signal. This offset is expected to switch sign near apocenter. This has been described as the \emph{slingshot effect} \citep{MandV:2007}, where the DM slows to a stop at apocenter, and the gas-core, which had fallen behind due to ram-pressure is accelerated toward the slowed subcluster and then shot past. This has been observed in Abell 168 \citep{Hallman:2004} and El Gordo \citep{Ng:2015}. In El Gordo, \citet{Ng:2015} showed that the radio relics are in agreement with the ram-pressure slingshot picture given the dynamical analysis. Thus, El Gordo is in the return phase of the merger ($\tau>\text{0.5}$). The Bullet Cluster and El Gordo thus may be viewed as two stages of the ram-pressure slingshot process, which occurs around $\tau=\text{0.5}$. To establish a relationship between the merger phase and the offset between the ICM and its host dark matter halo would require more clusters as well as hydrodynamic simulations. It is likely that this relationship would be most robust in clusters containing cool-core remnants like the Bullet Cluster and El Gordo. 

In ZwCl 0008, the X-ray surface brightness profile reveals a remnant gas-core associated with the west subcluster that has a trailing wake feature. This implies that the gas core is still outbound, and the fact that the gas-core remnant is still trailing its host subcluster in the west could be indicative of a merger that is still outbound in general, but the dark matter halo could have already reached apocenter, which would mean the ICM is undergoing the ram-pressure slingshot in the observed state. Unfortunately, the uncertainty in the lensing mass location and the small offset between the galaxy distribution and the ICM inhibits our ability to constrain the phase of the merger based on the slingshot effect. 

It is noteworthy that most time in spent near apocenter in a bimodal merger, thus it is possible that the seemingly discrepant results regarding the phase of ZwCl 0008 point to the merger's temporal proximity to apocenter. This is substantiated by the small line of sight velocity difference between the two subclusters, which could imply that the merger is observed near apocenter.

\subsection{Summary}

We have presented a rich dataset including X-ray, radio, and optical photometry and spectroscopy. ZwCl 0008 is composed of two subclusters merging along an east--west axis that is collinear with the two subclusters, the ``bullet''-like ICM, and two radio relics. We find the east subcluster to be more massive than the west subcluster. The X-ray surface brightness displays a beautiful gas remnant and wake feature associated with the west subcluster. The trailing wake indicates that the gas-core remnant is moving to the west in the observed state; however, we are unable to confidently state that the dark matter halos are still traveling outwards or they have begun returning post-apocenter. 

The radio relic polarization indicates the merger is occurring within $\sim\text{40}^{\circ}$ of the plane of the sky. With the above results as input into a dynamical Monte Carlo analysis, the cluster is found to have merged with a velocity of $\text{1800}^{+\text{400}}_{-\text{300}}\,\text{km}\,\text{s}^{-\text{1}}$. Pericenter occurred $\text{0.76}^{+\text{0.24}}_{-\text{0.27}}\,\text{Gyr}$ or $\text{1.3}^{+\text{0.90}}_{-\text{0.35}}\,\text{Gyr}$ ago in the outbound and return scenarios, respectively. While we are unable to select between the outbound and return scenarios, the positions of the gas and BCG may suggest that ZwCl 0008 in an intermediate stage of the ram-pressure slingshot. We find substantial offsets between the total mass and galaxy distributions in each subcluster; however, these offsets are statistically insignificant given the low lensing signal in our data. Finally, the similarities to the Bullet Cluster could prove to be valuable for tests of alternative DM models. Such tests will be more robust if we have many Bullet-like clusters at a variety of phases and merger speeds, especially if they are transverse such as ZwCl 0008.

\section{Acknowledgments}
We would like to thank the broader membership of the Merging Cluster Collaboration for their continual development of the science motivating this work.

This material is based upon work supported by the National Science Foundation under Grant No. (1518246).

This material is based upon work supported by STSci grant HST-GO-13343.001-A.

Support for this work was provided by the National Aeronautics and Space Administration through Chandra Award Number GO5-14130X issued by the Chandra X-ray Observatory Center, which is operated by the Smithsonian Astrophysical Observatory for and on behalf of the National Aeronautics Space Administration under contract NAS8-03060.

Part of this was work performed under the auspices of the U.S. DOE by LLNL under Contract DE-AC52-07NA27344.

RJW is supported by a Clay Fellowship awarded by the Harvard-Smithsonian Center for Astrophysics.

This research has made use of the NASA/IPAC Extragalactic Database (NED) which is operated by the Jet Propulsion Laboratory, California Institute of Technology, under contract with the National Aeronautics and Space Administration.

This research has made use of NASA's Astrophysics Data System.

This research made use of Montage. It is funded by the National Science Foundation under Grant Number ACI-1440620, and was previously funded by the National Aeronautics and Space Administration's Earth Science Technology Office, Computation Technologies Project, under Cooperative Agreement Number NCC5-626 between NASA and the California Institute of Technology.

This research made use of APLpy, an open-source plotting package for Python hosted at {\tt http://aplpy.github.com}.

{\it Facilities:} \facility{Keck II (DEIMOS)}, \facility{Subaru (SuprimeCam)}, \facility{Chandra (ACIS)}, \facility{INT (WFC)}, \facility{JVLA (C/D)}, \facility{Hubble Space Telescope (ACS, WFC3)}.

\bibliographystyle{aasjournal.bst}
\bibliography{mcc}

\begin{thebibliography}{}
\expandafter\ifx\csname natexlab\endcsname\relax\def\natexlab#1{#1}\fi
\providecommand{\url}[1]{\href{#1}{#1}}

\bibitem[{{Arnaud}(1996)}]{Arnaud:2006}
{Arnaud}, K.~A. 1996, in Astronomical Society of the Pacific Conference Series,
  Vol. 101, Astronomical Data Analysis Software and Systems V, ed. G.~H.
  {Jacoby} \& J.~{Barnes}, 17

\bibitem[{{Bahcall} \& {Kulier}(2014)}]{Bahcall:2014}
{Bahcall}, N.~A., \& {Kulier}, A. 2014, \mnras, 439, 2505

\bibitem[{{Bartelmann} \& {Schneider}(2001)}]{Bartelmann:2001}
{Bartelmann}, M., \& {Schneider}, P. 2001, \physrep, 340, 291

\bibitem[{{Beers} {et~al.}(1990){Beers}, {Flynn}, \& {Gebhardt}}]{Beers1990}
{Beers}, T.~C., {Flynn}, K., \& {Gebhardt}, K. 1990, \aj, 100, 32

\bibitem[{{Bertin}(2006)}]{SCAMP}
{Bertin}, E. 2006, in Astronomical Society of the Pacific Conference Series,
  Vol. 351, Astronomical Data Analysis Software and Systems XV, ed.
  C.~{Gabriel}, C.~{Arviset}, D.~{Ponz}, \& S.~{Enrique}, 112

\bibitem[{{Bertin} \& {Arnouts}(1996)}]{sextractor}
{Bertin}, E., \& {Arnouts}, S. 1996, \aaps, 117, 393

\bibitem[{{Bertin} {et~al.}(2002){Bertin}, {Mellier}, {Radovich}, {Missonnier},
  {Didelon}, \& {Morin}}]{SWARP}
{Bertin}, E., {Mellier}, Y., {Radovich}, M., {et~al.} 2002, in Astronomical
  Society of the Pacific Conference Series, Vol. 281, Astronomical Data
  Analysis Software and Systems XI, ed. D.~A. {Bohlender}, D.~{Durand}, \&
  T.~H. {Handley}, 228

\bibitem[{{Bonafede} {et~al.}(2010){Bonafede}, {Feretti}, {Murgia}, {Govoni},
  {Giovannini}, \& {Vacca}}]{Bonafede:2010}
{Bonafede}, A., {Feretti}, L., {Murgia}, M., {et~al.} 2010, ArXiv e-prints,
  arXiv:1009.1233

\bibitem[{{Briggs}(1995)}]{briggs_phd}
{Briggs}, D. 1995, PhD thesis, New Mexico Institute of Mining and Technology

\bibitem[{{Buton} {et~al.}(2013){Buton}, {Copin}, {Aldering}, {Antilogus},
  {Aragon}, {Bailey}, {Baltay}, {Bongard}, {Canto}, {Cellier-Holzem},
  {Childress}, {Chotard}, {Fakhouri}, {Gangler}, {Guy}, {Hsiao}, {Kerschhaggl},
  {Kowalski}, {Loken}, {Nugent}, {Paech}, {Pain}, {P{\'e}contal}, {Pereira},
  {Perlmutter}, {Rabinowitz}, {Rigault}, {Runge}, {Scalzo}, {Smadja}, {Tao},
  {Thomas}, {Weaver}, {Wu}, \& {Nearby SuperNova Factory}}]{Buton:2013}
{Buton}, C., {Copin}, Y., {Aldering}, G., {et~al.} 2013, \aap, 549, A8

\bibitem[{{Carlberg}(1994)}]{Carlberg:1994}
{Carlberg}, R.~G. 1994, \apj, 433, 468

\bibitem[{{Cavaliere} \& {Fusco-Femiano}(1976)}]{Cavaliere:1976}
{Cavaliere}, A., \& {Fusco-Femiano}, R. 1976, \aap, 49, 137

\bibitem[{{Clowe} {et~al.}(2006){Clowe}, {Bradac}, {Gonzalez}, {Markevitch},
  {Randall}, {Jones}, \& {Zaritsky}}]{Clowe06}
{Clowe}, D., {Bradac}, M., {Gonzalez}, A., {et~al.} 2006, \apjl, 648, L109

\bibitem[{{Cooper} {et~al.}(2012){Cooper}, {Newman}, {Davis}, {Finkbeiner}, \&
  {Gerke}}]{spec2d}
{Cooper}, M.~C., {Newman}, J.~A., {Davis}, M., {Finkbeiner}, D.~P., \& {Gerke},
  B.~F. 2012, {spec2d: DEEP2 DEIMOS Spectral Pipeline}, Astrophysics Source
  Code Library, , , ascl:1203.003

\bibitem[{{Cornwell} {et~al.}(2005){Cornwell}, {Golap}, \&
  {Bhatnagar}}]{Cornwell:2005}
{Cornwell}, T.~J., {Golap}, K., \& {Bhatnagar}, S. 2005, in Astronomical
  Society of the Pacific Conference Series, Vol. 347, Astronomical Data
  Analysis Software and Systems XIV, ed. P.~{Shopbell}, M.~{Britton}, \&
  R.~{Ebert}, 86

\bibitem[{{Cornwell} {et~al.}(2008){Cornwell}, {Golap}, \&
  {Bhatnagar}}]{Cornwell:2008}
{Cornwell}, T.~J., {Golap}, K., \& {Bhatnagar}, S. 2008, IEEE Journal of
  Selected Topics in Signal Processing, 2, 647

\bibitem[{{Dahlen} {et~al.}(2010){Dahlen}, {Mobasher}, {Dickinson}, {Ferguson},
  {Giavalisco}, {Grogin}, {Guo}, {Koekemoer}, {Lee}, {Lee}, {Nonino}, {Riess},
  \& {Salimbeni}}]{Dahlen:2010}
{Dahlen}, T., {Mobasher}, B., {Dickinson}, M., {et~al.} 2010, \apj, 724, 425

\bibitem[{{Dawson}(2013)}]{Dawson:2012}
{Dawson}, W.~A. 2013, \apj, 772, 131

\bibitem[{{Dawson}(2014)}]{MCMAC}
---. 2014, {MCMAC: Monte Carlo Merger Analysis Code}, Astrophysics Source Code
  Library, , , ascl:1407.004

\bibitem[{{Dawson} {et~al.}(2015){Dawson}, {Jee}, {Stroe}, {Ng}, {Golovich},
  {Wittman}, {Sobral}, {Br{\"u}ggen}, {R{\"o}ttgering}, \& {van
  Weeren}}]{Dawson:2014}
{Dawson}, W.~A., {Jee}, M.~J., {Stroe}, A., {et~al.} 2015, \apj, 805, 143

\bibitem[{{de Gasperin} {et~al.}(2015){de Gasperin}, {Intema}, {van Weeren},
  {Dawson}, {Golovich}, {Wittman}, {Bonafede}, \&
  {Br{\"u}ggen}}]{deGasperin:2015}
{de Gasperin}, F., {Intema}, H.~T., {van Weeren}, R.~J., {et~al.} 2015, \mnras,
  453, 3483

\bibitem[{{de Gasperin} {et~al.}(2014){de Gasperin}, {van Weeren},
  {Br{\"u}ggen}, {Vazza}, {Bonafede}, \& {Intema}}]{deGasperin:2014}
{de Gasperin}, F., {van Weeren}, R.~J., {Br{\"u}ggen}, M., {et~al.} 2014,
  \mnras, 444, 3130

\bibitem[{{Duffy} {et~al.}(2008){Duffy}, {Schaye}, {Kay}, \& {Dalla
  Vecchia}}]{Duffy}
{Duffy}, A.~R., {Schaye}, J., {Kay}, S.~T., \& {Dalla Vecchia}, C. 2008,
  \mnras, 390, L64

\bibitem[{{Eckert} {et~al.}(2011){Eckert}, {Molendi}, \&
  {Paltani}}]{Eckert:2011}
{Eckert}, D., {Molendi}, S., \& {Paltani}, S. 2011, \aap, 526, A79

\bibitem[{{Ensslin} {et~al.}(1998){Ensslin}, {Biermann}, {Klein}, \&
  {Kohle}}]{ensslin1998}
{Ensslin}, T.~A., {Biermann}, P.~L., {Klein}, U., \& {Kohle}, S. 1998, \aap,
  332, 395

\bibitem[{{Evrard} {et~al.}(2008){Evrard}, {Bialek}, {Busha}, {White}, {Habib},
  {Heitmann}, {Warren}, {Rasia}, {Tormen}, {Moscardini}, {Power}, {Jenkins},
  {Gao}, {Frenk}, {Springel}, {White}, \& {Diemand}}]{Evrard:2008}
{Evrard}, A.~E., {Bialek}, J., {Busha}, M., {et~al.} 2008, \apj, 672, 122

\bibitem[{{Faber} {et~al.}(2003){Faber}, {Phillips}, {Kibrick}, {Alcott},
  {Allen}, {Burrous}, {Cantrall}, {Clarke}, {Coil}, {Cowley}, {Davis}, {Deich},
  {Dietsch}, {Gilmore}, {Harper}, {Hilyard}, {Lewis}, {McVeigh}, {Newman},
  {Osborne}, {Schiavon}, {Stover}, {Tucker}, {Wallace}, {Wei}, {Wirth}, \&
  {Wright}}]{DEIMOS}
{Faber}, S.~M., {Phillips}, A.~C., {Kibrick}, R.~I., {et~al.} 2003, in
  \procspie, Vol. 4841, Instrument Design and Performance for Optical/Infrared
  Ground-based Telescopes, ed. M.~{Iye} \& A.~F.~M. {Moorwood}, 1657--1669

\bibitem[{{Feretti} {et~al.}(2012){Feretti}, {Giovannini}, {Govoni}, \&
  {Murgia}}]{Feretti:2012}
{Feretti}, L., {Giovannini}, G., {Govoni}, F., \& {Murgia}, M. 2012, \aapr, 20,
  54

\bibitem[{{Filippenko}(1982)}]{Filippenko}
{Filippenko}, A.~V. 1982, \pasp, 94, 715

\bibitem[{{Finoguenov} {et~al.}(2001){Finoguenov}, {Reiprich}, \&
  {B{\"o}hringer}}]{Finoguenov:2001}
{Finoguenov}, A., {Reiprich}, T.~H., \& {B{\"o}hringer}, H. 2001, \aap, 368,
  749

\bibitem[{{Finoguenov} {et~al.}(2010){Finoguenov}, {Sarazin}, {Nakazawa},
  {Wik}, \& {Clarke}}]{Finoguenov:2010}
{Finoguenov}, A., {Sarazin}, C.~L., {Nakazawa}, K., {Wik}, D.~R., \& {Clarke},
  T.~E. 2010, \apj, 715, 1143

\bibitem[{{Giavalisco} {et~al.}(2004){Giavalisco}, {Ferguson}, {Koekemoer},
  {Dickinson}, {Alexander}, {Bauer}, {Bergeron}, {Biagetti}, {Brandt},
  {Casertano}, {Cesarsky}, {Chatzichristou}, {Conselice}, {Cristiani}, {Da
  Costa}, {Dahlen}, {de Mello}, {Eisenhardt}, {Erben}, {Fall}, {Fassnacht},
  {Fosbury}, {Fruchter}, {Gardner}, {Grogin}, {Hook}, {Hornschemeier}, {Idzi},
  {Jogee}, {Kretchmer}, {Laidler}, {Lee}, {Livio}, {Lucas}, {Madau},
  {Mobasher}, {Moustakas}, {Nonino}, {Padovani}, {Papovich}, {Park},
  {Ravindranath}, {Renzini}, {Richardson}, {Riess}, {Rosati}, {Schirmer},
  {Schreier}, {Somerville}, {Spinrad}, {Stern}, {Stiavelli}, {Strolger},
  {Urry}, {Vandame}, {Williams}, \& {Wolf}}]{GOODS}
{Giavalisco}, M., {Ferguson}, H.~C., {Koekemoer}, A.~M., {et~al.} 2004, \apjl,
  600, L93

\bibitem[{{Golovich} {et~al.}(2016){Golovich}, {Dawson}, {Wittman}, {Ogrean},
  {van Weeren}, \& {Bonafede}}]{Golovich:2016}
{Golovich}, N., {Dawson}, W.~A., {Wittman}, D., {et~al.} 2016, \apj, 831, 110

\bibitem[{{Hallman} \& {Markevitch}(2004)}]{Hallman:2004}
{Hallman}, E.~J., \& {Markevitch}, M. 2004, \apjl, 610, L81

\bibitem[{{Hoekstra}(2013)}]{Hoekstra:2013}
{Hoekstra}, H. 2013, ArXiv e-prints, arXiv:1312.5981

\bibitem[{{Jee} {et~al.}(2016){Jee}, {Dawson}, {Stroe}, {Wittman}, {van
  Weeren}, {Br{\"u}ggen}, {Brada{\v c}}, \& {R{\"o}ttgering}}]{Jee:2016}
{Jee}, M.~J., {Dawson}, W.~A., {Stroe}, A., {et~al.} 2016, \apj, 817, 179

\bibitem[{{Jee} {et~al.}(2014){Jee}, {Hughes}, {Menanteau}, {Sif{\'o}n},
  {Mandelbaum}, {Barrientos}, {Infante}, \& {Ng}}]{Jee:2014}
{Jee}, M.~J., {Hughes}, J.~P., {Menanteau}, F., {et~al.} 2014, \apj, 785, 20

\bibitem[{{Jee} {et~al.}(2007){Jee}, {Ford}, {Illingworth}, {White},
  {Broadhurst}, {Coe}, {Meurer}, {van der Wel}, {Ben{\'{\i}}tez}, {Blakeslee},
  {Bouwens}, {Bradley}, {Demarco}, {Homeier}, {Martel}, \& {Mei}}]{Jee07}
{Jee}, M.~J., {Ford}, H.~C., {Illingworth}, G.~D., {et~al.} 2007, \apj, 661,
  728

\bibitem[{{Jee} {et~al.}(2015){Jee}, {Stroe}, {Dawson}, {Wittman}, {Hoekstra},
  {Br{\"u}ggen}, {R{\"o}ttgering}, {Sobral}, \& {van Weeren}}]{Jee:2015}
{Jee}, M.~J., {Stroe}, A., {Dawson}, W., {et~al.} 2015, \apj, 802, 46

\bibitem[{{Kalberla} {et~al.}(2005){Kalberla}, {Burton}, {Hartmann}, {Arnal},
  {Bajaja}, {Morras}, \& {P{\"o}ppel}}]{Kalberla}
{Kalberla}, P.~M.~W., {Burton}, W.~B., {Hartmann}, D., {et~al.} 2005, \aap,
  440, 775

\bibitem[{{Kang} {et~al.}(2012){Kang}, {Ryu}, \& {Jones}}]{Kang:2012}
{Kang}, H., {Ryu}, D., \& {Jones}, T.~W. 2012, \apj, 756, 97

\bibitem[{Kass \& Raftery(1995)}]{Kass:1995}
Kass, R.~E., \& Raftery, A.~E. 1995, Journal of the American Statistical
  Association, 90, 773

\bibitem[{{Kierdorf} {et~al.}(2016){Kierdorf}, {Beck}, {Hoeft}, {Klein}, {van
  Weeren}, {Forman}, \& {Jones}}]{Kierdorf:2016}
{Kierdorf}, M., {Beck}, R., {Hoeft}, M., {et~al.} 2016, ArXiv e-prints,
  arXiv:1612.01764

\bibitem[{{Kim} {et~al.}(2016){Kim}, {Peter}, \& {Wittman}}]{Kim:2016}
{Kim}, S.~Y., {Peter}, A.~H.~G., \& {Wittman}, D. 2016, ArXiv e-prints,
  arXiv:1608.08630

\bibitem[{Koekemoer {et~al.}(2003)Koekemoer, Fruchter, Hook, \&
  Hack}]{multidrizzle}
Koekemoer, A.~M., Fruchter, A.~S., Hook, R., \& Hack, W. 2003, in HST
  Calibration Workshop: Hubble after the Installation of the ACS and the NICMOS
  Cooling System, Vol.~1, 337

\bibitem[{{Macario} {et~al.}(2011){Macario}, {Markevitch}, {Giacintucci},
  {Brunetti}, {Venturi}, \& {Murray}}]{Macario:2011}
{Macario}, G., {Markevitch}, M., {Giacintucci}, S., {et~al.} 2011, \apj, 728,
  82

\bibitem[{{Markevitch}(2006)}]{markevitch05}
{Markevitch}, M. 2006, in ESA Special Publication, Vol. 604, The X-ray Universe
  2005, ed. A.~{Wilson}, 723

\bibitem[{{Markevitch} {et~al.}(2004){Markevitch}, {Gonzalez}, {Clowe},
  {Vikhlinin}, {Forman}, {Jones}, {Murray}, \& {Tucker}}]{Markevitch04}
{Markevitch}, M., {Gonzalez}, A.~H., {Clowe}, D., {et~al.} 2004, \apj, 606, 819

\bibitem[{{Markevitch} \& {Vikhlinin}(2007)}]{MandV:2007}
{Markevitch}, M., \& {Vikhlinin}, A. 2007, \physrep, 443, 1

\bibitem[{{McMullin} {et~al.}(2007){McMullin}, {Waters}, {Schiebel}, {Young},
  \& {Golap}}]{McMullin:2007}
{McMullin}, J.~P., {Waters}, B., {Schiebel}, D., {Young}, W., \& {Golap}, K.
  2007, in Astronomical Society of the Pacific Conference Series, Vol. 376,
  Astronomical Data Analysis Software and Systems XVI, ed. R.~A. {Shaw},
  F.~{Hill}, \& D.~J. {Bell}, 127

\bibitem[{{Merten} {et~al.}(2011){Merten}, {Coe}, {Dupke}, {Massey}, {Zitrin},
  {Cypriano}, {Okabe}, {Frye}, {Braglia}, {Jim{\'e}nez-Teja}, {Ben{\'{\i}}tez},
  {Broadhurst}, {Rhodes}, {Meneghetti}, {Moustakas}, {Sodr{\'e}}, {Krick}, \&
  {Bregman}}]{Merten:2011}
{Merten}, J., {Coe}, D., {Dupke}, R., {et~al.} 2011, \mnras, 417, 333

\bibitem[{{Mohan} \& {Rafferty}(2015)}]{Mohan:2015}
{Mohan}, N., \& {Rafferty}, D. 2015, {PyBDSM: Python Blob Detection and Source
  Measurement}, Astrophysics Source Code Library, , , ascl:1502.007

\bibitem[{{Monteiro-Oliveira} {et~al.}(2016){Monteiro-Oliveira}, {Cypriano},
  {Machado}, {Lima-Neto}, {Ribeiro}, {Sodr{\'e}}, \& {Dupke}}]{Monteiro:2016}
{Monteiro-Oliveira}, R., {Cypriano}, E.~S., {Machado}, R.~E.~G., {et~al.} 2016,
  ArXiv e-prints, arXiv:1605.07595

\bibitem[{{Newman} {et~al.}(2013){Newman}, {Cooper}, {Davis}, {Faber}, {Coil},
  {Guhathakurta}, {Koo}, {Phillips}, {Conroy}, {Dutton}, {Finkbeiner}, {Gerke},
  {Rosario}, {Weiner}, {Willmer}, {Yan}, {Harker}, {Kassin}, {Konidaris},
  {Lai}, {Madgwick}, {Noeske}, {Wirth}, {Connolly}, {Kaiser}, {Kirby},
  {Lemaux}, {Lin}, {Lotz}, {Luppino}, {Marinoni}, {Matthews}, {Metevier}, \&
  {Schiavon}}]{DEEP2:2013}
{Newman}, J.~A., {Cooper}, M.~C., {Davis}, M., {et~al.} 2013, \apjs, 208, 5

\bibitem[{{Ng} {et~al.}(2015){Ng}, {Dawson}, {Wittman}, {Jee}, {Hughes},
  {Menanteau}, \& {Sif{\'o}n}}]{Ng:2015}
{Ng}, K.~Y., {Dawson}, W.~A., {Wittman}, D., {et~al.} 2015, \mnras, 453, 1531

\bibitem[{{Offringa} {et~al.}(2010){Offringa}, {de Bruyn}, {Biehl}, {Zaroubi},
  {Bernardi}, \& {Pandey}}]{Offringa:2010}
{Offringa}, A.~R., {de Bruyn}, A.~G., {Biehl}, M., {et~al.} 2010, \mnras, 405,
  155

\bibitem[{{Offringa} {et~al.}(2014){Offringa}, {McKinley}, {Hurley-Walker},
  {Briggs}, {Wayth}, {Kaplan}, {Bell}, {Feng}, {Neben}, {Hughes}, {Rhee},
  {Murphy}, {Bhat}, {Bernardi}, {Bowman}, {Cappallo}, {Corey}, {Deshpande},
  {Emrich}, {Ewall-Wice}, {Gaensler}, {Goeke}, {Greenhill}, {Hazelton},
  {Hindson}, {Johnston-Hollitt}, {Jacobs}, {Kasper}, {Kratzenberg}, {Lenc},
  {Lonsdale}, {Lynch}, {McWhirter}, {Mitchell}, {Morales}, {Morgan},
  {Kudryavtseva}, {Oberoi}, {Ord}, {Pindor}, {Procopio}, {Prabu}, {Riding},
  {Roshi}, {Shankar}, {Srivani}, {Subrahmanyan}, {Tingay}, {Waterson},
  {Webster}, {Whitney}, {Williams}, \& {Williams}}]{Offringa:2014}
{Offringa}, A.~R., {McKinley}, B., {Hurley-Walker}, N., {et~al.} 2014, \mnras,
  444, 606

\bibitem[{{Ogrean} {et~al.}(2013){Ogrean}, {Br{\"u}ggen}, {van Weeren},
  {R{\"o}ttgering}, {Croston}, \& {Hoeft}}]{Ogrean:2013c}
{Ogrean}, G.~A., {Br{\"u}ggen}, M., {van Weeren}, R.~J., {et~al.} 2013, \mnras,
  433, 812

\bibitem[{{Ouchi} {et~al.}(2004){Ouchi}, {Shimasaku}, {Okamura}, {Furusawa},
  {Kashikawa}, {Ota}, {Doi}, {Hamabe}, {Kimura}, {Komiyama}, {Miyazaki},
  {Miyazaki}, {Nakata}, {Sekiguchi}, {Yagi}, \& {Yasuda}}]{Ouchi:2004}
{Ouchi}, M., {Shimasaku}, K., {Okamura}, S., {et~al.} 2004, \apj, 611, 660

\bibitem[{{Paul} {et~al.}(2011){Paul}, {Iapichino}, {Miniati}, {Bagchi}, \&
  {Mannheim}}]{Paul:2011}
{Paul}, S., {Iapichino}, L., {Miniati}, F., {Bagchi}, J., \& {Mannheim}, K.
  2011, \apj, 726, 17

\bibitem[{Pedregosa {et~al.}(2011)Pedregosa, Varoquaux, Gramfort, Michel,
  Thirion, Grisel, Blondel, Prettenhofer, Weiss, Dubourg, Vanderplas, Passos,
  Cournapeau, Brucher, Perrot, \& Duchesnay}]{sklearn}
Pedregosa, F., Varoquaux, G., Gramfort, A., {et~al.} 2011, Journal of Machine
  Learning Research, 12, 2825

\bibitem[{{Pinkney} {et~al.}(1996){Pinkney}, {Roettiger}, {Burns}, \&
  {Bird}}]{Pinkney:1996}
{Pinkney}, J., {Roettiger}, K., {Burns}, J.~O., \& {Bird}, C.~M. 1996, \apjs,
  104, 1

\bibitem[{{Planck Collaboration} {et~al.}(2015){Planck Collaboration}, {Ade},
  {Aghanim}, {Arnaud}, {Ashdown}, {Aumont}, {Baccigalupi}, {Banday},
  {Barreiro}, {Bartlett}, \& et~al.}]{PlanckMass}
{Planck Collaboration}, {Ade}, P.~A.~R., {Aghanim}, N., {et~al.} 2015, ArXiv
  e-prints, arXiv:1502.01597

\bibitem[{{Pratt} {et~al.}(2009){Pratt}, {Croston}, {Arnaud}, \&
  {B{\"o}hringer}}]{Pratt:2009}
{Pratt}, G.~W., {Croston}, J.~H., {Arnaud}, M., \& {B{\"o}hringer}, H. 2009,
  \aap, 498, 361

\bibitem[{{Randall} {et~al.}(2008){Randall}, {Markevitch}, {Clowe}, {Gonzalez},
  \& {Bradac}}]{randall2008}
{Randall}, S.~W., {Markevitch}, M., {Clowe}, D., {Gonzalez}, A.~H., \&
  {Bradac}, M. 2008, ApJ, 679, 1173

\bibitem[{{Rau} \& {Cornwell}(2011)}]{Rau:2011}
{Rau}, U., \& {Cornwell}, T.~J. 2011, \aap, 532, A71

\bibitem[{{Riseley} {et~al.}(2017){Riseley}, {Scaife}, {Wise}, \&
  {Clarke}}]{Riseley:2017}
{Riseley}, C.~J., {Scaife}, A.~M.~M., {Wise}, M.~W., \& {Clarke}, A.~O. 2017,
  \aap, 597, A96

\bibitem[{{Schlafly} \& {Finkbeiner}(2011)}]{Schlafly:2011}
{Schlafly}, E.~F., \& {Finkbeiner}, D.~P. 2011, \apj, 737, 103

\bibitem[{{Shimwell} {et~al.}(2015){Shimwell}, {Markevitch}, {Brown},
  {Feretti}, {Gaensler}, {Johnston-Hollitt}, {Lage}, \&
  {Srinivasan}}]{Shimwell:2015}
{Shimwell}, T.~W., {Markevitch}, M., {Brown}, S., {et~al.} 2015, \mnras, 449,
  1486

\bibitem[{{Shimwell} {et~al.}(2016){Shimwell}, {Luckin}, {Br{\"u}ggen},
  {Brunetti}, {Intema}, {Owers}, {R{\"o}ttgering}, {Stroe}, {van Weeren},
  {Williams}, {Cassano}, {de Gasperin}, {Heald}, {Hoang}, {Hardcastle},
  {Sridhar}, {Sabater}, {Best}, {Bonafede}, {Chy{\.z}y}, {En{\ss}lin},
  {Ferrari}, {Haverkorn}, {Hoeft}, {Horellou}, {McKean}, {Morabito},
  {Orr{\`u}}, {Pizzo}, {Retana-Montenegro}, \& {White}}]{Shimwell:2016}
{Shimwell}, T.~W., {Luckin}, J., {Br{\"u}ggen}, M., {et~al.} 2016, \mnras, 459,
  277

\bibitem[{{Skillman} {et~al.}(2013){Skillman}, {Xu}, {Hallman}, {O'Shea},
  {Burns}, {Li}, {Collins}, \& {Norman}}]{Skillman:2013}
{Skillman}, S.~W., {Xu}, H., {Hallman}, E.~J., {et~al.} 2013, \apj, 765, 21

\bibitem[{{Skrutskie} {et~al.}(2006){Skrutskie}, {Cutri}, {Stiening},
  {Weinberg}, {Schneider}, {Carpenter}, {Beichman}, {Capps}, {Chester},
  {Elias}, {Huchra}, {Liebert}, {Lonsdale}, {Monet}, {Price}, {Seitzer},
  {Jarrett}, {Kirkpatrick}, {Gizis}, {Howard}, {Evans}, {Fowler}, {Fullmer},
  {Hurt}, {Light}, {Kopan}, {Marsh}, {McCallon}, {Tam}, {Van Dyk}, \&
  {Wheelock}}]{2mass}
{Skrutskie}, M.~F., {Cutri}, R.~M., {Stiening}, R., {et~al.} 2006, \aj, 131,
  1163

\bibitem[{{Sobral} {et~al.}(2015){Sobral}, {Stroe}, {Dawson}, {Wittman}, {Jee},
  {R{\"o}ttgering}, {van Weeren}, \& {Br{\"u}ggen}}]{Sobral:2015}
{Sobral}, D., {Stroe}, A., {Dawson}, W.~A., {et~al.} 2015, /mnras, 450, 630

\bibitem[{{Springel} \& {Farrar}(2007)}]{springel2007}
{Springel}, V., \& {Farrar}, G.~R. 2007, MNRAS, 380, 911

\bibitem[{{Takizawa}(1999)}]{Takizawa:1999}
{Takizawa}, M. 1999, \apj, 520, 514

\bibitem[{{Takizawa} {et~al.}(2010){Takizawa}, {Nagino}, \&
  {Matsushita}}]{Takizawa:2010}
{Takizawa}, M., {Nagino}, R., \& {Matsushita}, K. 2010, \pasj, 62, 951

\bibitem[{{Taylor} {et~al.}(2009){Taylor}, {Stil}, \& {Sunstrum}}]{Taylor:2009}
{Taylor}, A.~R., {Stil}, J.~M., \& {Sunstrum}, C. 2009, \apj, 702, 1230

\bibitem[{{\'U}beda \& Anderson(2012)}]{HSTCTI}
{\'U}beda, L., \& Anderson, J. 2012, ACS Instrument Science Report, 3, 2012

\bibitem[{{van Weeren} {et~al.}(2011{\natexlab{a}}){van Weeren}, {Br{\"u}ggen},
  {R{\"o}ttgering}, \& {Hoeft}}]{vanWeeren:2011b}
{van Weeren}, R.~J., {Br{\"u}ggen}, M., {R{\"o}ttgering}, H.~J.~A., \& {Hoeft},
  M. 2011{\natexlab{a}}, \mnras, 418, 230

\bibitem[{{van Weeren} {et~al.}(2011{\natexlab{b}}){van Weeren}, {Br{\"u}ggen},
  {R{\"o}ttgering}, {Hoeft}, {Nuza}, \& {Intema}}]{vanWeeren:2011}
{van Weeren}, R.~J., {Br{\"u}ggen}, M., {R{\"o}ttgering}, H.~J.~A., {et~al.}
  2011{\natexlab{b}}, \aap, 533, A35

\bibitem[{{van Weeren} {et~al.}(2011{\natexlab{c}}){van Weeren}, {Hoeft},
  {R{\"o}ttgering}, {Br{\"u}ggen}, {Intema}, \& {van Velzen}}]{vanWeeren2011}
{van Weeren}, R.~J., {Hoeft}, M., {R{\"o}ttgering}, H.~J.~A., {et~al.}
  2011{\natexlab{c}}, \aap, 528, A38

\bibitem[{{van Weeren} {et~al.}(2010){van Weeren}, {R{\"o}ttgering},
  {Br{\"u}ggen}, \& {Hoeft}}]{vanweeren2010}
{van Weeren}, R.~J., {R{\"o}ttgering}, H.~J.~A., {Br{\"u}ggen}, M., \& {Hoeft},
  M. 2010, Science, 330, 347

\bibitem[{{van Weeren} {et~al.}(2016){van Weeren}, {Ogrean}, {Jones}, {Forman},
  {Andrade-Santos}, {Bonafede}, {Br{\"u}ggen}, {Bulbul}, {Clarke}, {Churazov},
  {David}, {Dawson}, {Donahue}, {Goulding}, {Kraft}, {Mason}, {Merten},
  {Mroczkowski}, {Murray}, {Nulsen}, {Rosati}, {Roediger}, {Randall}, {Sayers},
  {Umetsu}, {Vikhlinin}, \& {Zitrin}}]{vanWeeren:2016}
{van Weeren}, R.~J., {Ogrean}, G.~A., {Jones}, C., {et~al.} 2016, \apj, 817, 98

\bibitem[{{van Weeren} {et~al.}(2017){van Weeren}, {Andrade-Santos}, {Dawson},
  {Golovich}, {Lal}, {Kang}, {Ryu}, {Br{\`i}ggen}, {Ogrean}, {Forman}, {Jones},
  {Placco}, {Santucci}, {Wittman}, {Jee}, {Kraft}, {Sobral}, {Stroe}, \&
  {Fogarty}}]{vanWeeren:2017}
{van Weeren}, R.~J., {Andrade-Santos}, F., {Dawson}, W.~A., {et~al.} 2017,
  Nature Astronomy, 1, 0005

\bibitem[{{Vikhlinin} {et~al.}(2005){Vikhlinin}, {Markevitch}, {Murray},
  {Jones}, {Forman}, \& {Van Speybroeck}}]{Vikhlinin2005}
{Vikhlinin}, A., {Markevitch}, M., {Murray}, S.~S., {et~al.} 2005, \apj, 628,
  655

\bibitem[{{Vogelsberger} {et~al.}(2014){Vogelsberger}, {Genel}, {Springel},
  {Torrey}, {Sijacki}, {Xu}, {Snyder}, {Nelson}, \& {Hernquist}}]{Illustris}
{Vogelsberger}, M., {Genel}, S., {Springel}, V., {et~al.} 2014, \mnras, 444,
  1518

\bibitem[{{Voges} {et~al.}(1999){Voges}, {Aschenbach}, {Boller},
  {Br{\"a}uninger}, {Briel}, {Burkert}, {Dennerl}, {Englhauser}, {Gruber},
  {Haberl}, {Hartner}, {Hasinger}, {K{\"u}rster}, {Pfeffermann}, {Pietsch},
  {Predehl}, {Rosso}, {Schmitt}, {Tr{\"u}mper}, \& {Zimmermann}}]{RASS}
{Voges}, W., {Aschenbach}, B., {Boller}, T., {et~al.} 1999, \aap, 349, 389

\bibitem[{{Wittman} {et~al.}(2006){Wittman}, {Dell'Antonio}, {Hughes},
  {Margoniner}, {Tyson}, {Cohen}, \& {Norman}}]{Wittman06}
{Wittman}, D., {Dell'Antonio}, I.~P., {Hughes}, J.~P., {et~al.} 2006, \apj,
  643, 128

\bibitem[{{Zhang} {et~al.}(2010){Zhang}, {Okabe}, {Finoguenov}, {Smith},
  {Piffaretti}, {Valdarnini}, {Babul}, {Evrard}, {Mazzotta}, {Sanderson}, \&
  {Marrone}}]{Zhang:2010}
{Zhang}, Y.-Y., {Okabe}, N., {Finoguenov}, A., {et~al.} 2010, \apj, 711, 1033

\end{thebibliography}
\end{document}